\documentclass[twocolumn]{aastex631}

\usepackage{amsmath}
\newcommand{\hst}{\textit{HST}}

\newcommand {\A}[1]{And~{\sc #1}}
\usepackage[caption=false]{subfig}

\shorttitle{Horizontal Branch Ages}
\shortauthors{Jennings et al.}
\begin{document}

\title{The Hubble Space Telescope Survey of M31 Satellite Galaxies. III. \\ Calibrating the Horizontal Branch as an Age Indicator for Nearby Galaxies}

\author{Connor Jennings}
\affiliation{Department of Astronomy, University of California, Berkeley, CA 94720-3411, USA}

\author[0000-0002-1445-4877]{Alessandro Savino}
\affiliation{Department of Astronomy, University of California, Berkeley, CA 94720-3411, USA}

\author[0000-0002-6442-6030]{Daniel R. Weisz}
\affiliation{Department of Astronomy, University of California, Berkeley, CA 94720-3411, USA}

\author[0000-0002-3204-1742]{Nitya Kallivayalil}
\affiliation{Department of Astronomy, The University of Virginia, 530 McCormick Road, Charlottesville, VA 22904, USA}

% \author[0000-0003-0603-8942]{Andrew Wetzel}
% \affiliation{Department of Physics and Astronomy, University of California, Davis, CA 95616, USA}

% \author{Jay Anderson}
% \affiliation{Space Telescope Science Institute, 3700 San Martin Drive, Baltimore, MD 21218, USA}

% \author[0000-0003-0715-2173]{Gurtina Besla}
% \affiliation{Department of Astronomy, University of Arizona, 933 North Cherry Avenue, Tucson, AZ 85721, USA}

% \author[0000-0002-9604-343X]{Michael Boylan-Kolchin}
% \affiliation{Department of Astronomy, The University of Texas at Austin, 2515 Speedway, Stop C1400, Austin, TX 78712, USA}

% \author{Thomas M. Brown}
% \affiliation{Space Telescope Science Institute, 3700 San Martin Drive, Baltimore, MD 21218, USA}

% \author{James S. Bullock}
% \affiliation{Department of Physics and Astronomy, University of California, Irvine, CA 92697 USA}

\author[0000-0003-0303-3855]{Andrew A. Cole}
\affiliation{School of Natural Sciences, University of Tasmania, Private Bag 37, Hobart, Tasmania 7001, Australia}

\author[0000-0002-1693-3265]{Michelle L.M. Collins}
\affiliation{Physics Department, University of Surrey, Guildford GU2 7XH, UK}

% \author[0000-0003-1371-6019]{M. C. Cooper}
% \affiliation{Department of Physics and Astronomy, University of California, Irvine, CA 92697 USA}

% \author[0000-0001-6146-2645]{Alis J. Deason}
% \affiliation{Institute for Computational Cosmology, Department of Physics, Durham University, Durham DH1 3LE, UK}

\author{Andrew Dolphin}
\affiliation{Raytheon Technologies, 1151 E. Hermans Road, Tucson, AZ 85756, USA}
\affiliation{Steward Observatory, University of Arizona, 933 N. Cherry Avenue, Tucson, AZ 85719, USA}

% \author{Aaron L. Dotter}
% \affiliation{Department of Physics and Astronomy, Dartmouth College, 6127 Wilder Laboratory, Hanover, NH 03755, USA}

% \author{Mark Fardal}
% % \affiliation{Space Telescope Science Institute, 3700 San Martin Drive, Baltimore, MD 21218, USA}
% \affiliation{Eureka Scientific, 2452 Delmer St., Suite 100, Oakland, CA 96402, USA}

\author{Annette M. N. Ferguson}
\affiliation{Institute for Astronomy, University of Edinburgh, Royal Observatory, Blackford Hill, Edinburgh, EH9 3HJ, UK}

% \author[0000-0002-3122-300X]{Tobias K. Fritz}
% \affiliation{ Department of Astronomy, University of Virginia, Charlottesville, 530 McCormick Road, VA 22904-4325, USA}

% \author[0000-0002-7007-9725]{Marla C. Geha}
% \affiliation{Department of Astronomy, Yale University, New Haven, CT 06520, USA}

\author[0000-0003-0394-8377]{Karoline M. Gilbert}
\affiliation{Department of Physics and Astronomy, Johns Hopkins University, Baltimore, MD 21218, USA}
\affiliation{Space Telescope Science Institute, 3700 San Martin Dr., Baltimore, MD 21218, USA}

% \affiliation{Space Telescope Science Institute, 3700 San Martin Drive, Baltimore, MD 21218, USA}
% \affiliation{The William H. Miller III Department of Physics \& Astronomy, Bloomberg Center for Physics and Astronomy, Johns Hopkins University, 3400 N. Charles Street, Baltimore, MD 21218}

\author{Puragra Guhathakurta}
\affiliation{UCO/Lick Observatory, Department of Astronomy \& Astrophysics, University of California Santa Cruz, 1156 High Street, Santa Cruz, California 95064, USA}

% \author{Rodrigo Ibata}
% \affiliation{Observatoire de Strasbourg, 11, rue de l’Universite, F-67000 Strasbourg, France}

% \author[0000-0002-2191-9038]{Michael J. Irwin}
% \affiliation{Institute of Astronomy, University of Cambridge, Cambridge CB3 0HA, UK}

% \author{Myoungwon Jeon}
% \affiliation{School of Space Research, Kyung Hee University, 1732 Deogyeong-daero, Yongin-si, Gyeonggi-do 17104, Republic of Korea}

\author{Evan N. Kirby}
\affiliation{Department of Physics, University of Notre Dame, Notre Dame, IN 46556, USA}

\author[0000-0003-3081-9319]{Geraint F. Lewis}
\affiliation{Sydney Institute for Astronomy, School of Physics, A28,
The University of Sydney, NSW 2006, Australia}

% \author[0000-0002-6529-8093]{Dougal Mackey}
% \affiliation{Research School of Astronomy and Astrophysics, Australian National
% University, Canberra 2611, ACT, Australia}

% \author{Steven R. Majewski}
% \affiliation{Department of Astronomy, University of Virginia, 530 McCormick Road, Charlottesville, VA 22904, USA}

\author[0000-0002-1349-202X]{Nicolas Martin}
\affiliation{Observatoire de Strasbourg, 11, rue de l’Universite, F-67000 Strasbourg, France}
\affiliation{Max-Planck-Institut fur Astronomie, K\"{o}nigstuhl 17, D-69117 Heidelberg, Germany}

% \author{Alan McConnachie}
% \affiliation{NRC Herzberg Astronomy and Astrophysics, 5071 West Saanich Road, Victoria, BC V9E 2E7, Canada}
% % \affiliation{Physics \& Astronomy Department, University of Victoria, 3800 Finnerty Road, Victoria, BC V8P 5C2 Canada}

% \author[0000-0002-9820-1219]{Ekta Patel}
% \affiliation{Department of Astronomy, University of California, Berkeley, Berkeley, CA, 94720, USA}

\author[0000-0003-0427-8387]{R. Michael Rich}
\affiliation{Department of Physics and Astronomy, UCLA, 430 Portola Plaza, Box 951547, Los Angeles, CA 90095-1547, USA}

% \author[0000-0002-4733-4994]{Joshua D. Simon}
% \affiliation{Observatories of the Carnegie Institution for Science, 813 Santa Barbara Street, Pasadena, CA 91101, USA}

\author[0000-0003-0605-8732]{Evan D. Skillman}
\affiliation{University of Minnesota, Minnesota Institute for Astrophysics, School of Physics and Astronomy, 116 Church Street, S.E., Minneapolis,
MN 55455, USA}

% \author[0000-0001-8368-0221]{Sangmo Tony Sohn}
% \affiliation{Space Telescope Science Institute, 3700 San Martin Drive, Baltimore, MD 21218, USA}

% \author{Erik J. Tollerud}
% \affiliation{Space Telescope Science Institute, 3700 San Martin Drive, Baltimore, MD 21218, USA}

\author[0000-0001-7827-7825]{Roeland P. van der Marel}
\affiliation{Space Telescope Science Institute, 3700 San Martin Drive, Baltimore, MD 21218, USA}
\affiliation{Center for Astrophysical Sciences,
The William H. Miller III Department of Physics \& Astronomy,
Johns Hopkins University, Baltimore, MD 21218, USA}

\author[0000-0003-1634-4644]{Jack T. Warfield}
\affiliation{Department of Astronomy, The University of Virginia, 530 McCormick Road, Charlottesville, VA
22904, USA}



\begin{abstract}

We present a new method for measuring the mean age of old/intermediate stellar populations in resolved, metal-poor ($\rm \langle[Fe/H]\rangle \lesssim -1.5$) galaxies using only the morphology of the horizontal branch (HB) and an estimate of the average metallicity. We calculate the ratio of blue-to-red HB stars and the mass-weighted mean ages of 27 M31 satellite galaxies that have star formation histories (SFHs) measured from Hubble Space Telescope-based color-magnitude diagrams (CMDs) that include the oldest Main Sequence Turn-off (MSTO) ages. We find a strong correlation between mean age, metallicity, and HB morphology, for stellar populations older than $\sim6$~Gyr.  The correlation allows us to predict a galaxy's mean age from its HB morphology to a precision of $\lesssim 1$~Gyr. We validate our method by recovering the correct ages of Local Group galaxies that have robust MSTO-based ages and are not in our calibration sample.  We also use our technique to measure the mean ages of isolated field galaxies KKR25 ($11.21^{+0.70}_{-0.65}$~Gyr) and VV124 ($11.03^{+0.73}_{-0.68}$~Gyr), which indicate that their main star formation episodes may have lasted several Gyr and support the picture that they achieved their early-type characteristics (e.g., low gas content, low star formation activity) in isolation and not through environment.  Because the HB is $\sim80\times$ brighter than the oldest MSTO, our method can provide precise characteristic ages of predominantly old galaxies at distances $\sim 9$ times farther. We provide our calibrations in commonly used HST/ACS filters.

%We present our calibration in F475W, F606W and F814W, which are commonly used HST filters for nearby galaxy studies. 

%We also use our technique to measure the mean ages of isolated field galaxies KKR25 ($10.84^{+0.84}_{-0.78}$~Gyr) and VV124 ($10.76^{+0.88}_{-0.81}$~Gyr), which indicate their main star formation episode lasted several Gyr and support the picture that they achieved their early-type characteristics in isolation.

\end{abstract}

\keywords{galaxies: dwarf -- Hertzsprung–Russell and C–M diagrams -- Local Group -- stars: horizontal-branch}

\section{Introduction} \label{sec:intro}
The color-magnitude diagrams (CMDs) of nearby galaxies provide detailed insight into their formation histories.  Several regions of a galaxy's CMD, such as the main sequence turnoff (MSTO), the sub-giant branch (SGB), and the red giant branch (RGB) are sensitive to age and/or metallicity, and provide the means to reconstruct detailed, `non-parametric' star formation and enrichment histories of galaxies over cosmic time \citep[e.g.,][]{Gallart05, Tolstoy09, Cignoni10}.  The exquisite angular resolution and sensitivity of the Hubble Space Telescope (\hst) has enabled the construction of CMDs and measurement of SFHs for hundreds of galaxies in and around the Local Group (LG), providing rich insight into galaxy evolution throughout the local Universe \citep[e.g.,][]{Dalcanton09, McQuinn10, Brown14, Weisz14, Gallart15, Skillman17}.

However, the accuracy of CMD-based star formation histories (SFHs) depends on the depth of the CMD.  The most robust SFHs are derived from CMDs that reach the oldest MSTO (oMSTO).  This faint feature ($M_{\rm oMSTO, V} \sim +4$) is sensitive to age and metallicity, while the relative simplicity of stellar physics for this phase of evolution translates to small systematic uncertainties on MSTO-based SFHs.  SFHs derived from CMDs that reach the oMSTO are considered to be accurate over all cosmic time \citep[e.g.,][]{Gallart05}.  

On the other hand, faintness and crowding effectively limit observations of the oMSTO to galaxies located within the LG, even with the resolving power of \hst.  Expanding galaxy ages and SFHs beyond the LG has required modeling more luminous, evolved phases of stellar evolution \citep[e.g., red giants, red clump;][]{Rejkuba05,Weisz11}, all of which are less sensitive to age than the MSTO and subject to much less certain physics. The result is that SFHs from shallower CMDs only provide coarse age resolution over cosmic time and systematics in the stellar libraries themselves can be the dominant source of uncertainty.

The horizontal branch (HB) offers a promising compromise between SFH fidelity and observational access. In the optical, it is 3-4 magnitudes brighter than the oMSTO, while its sensitivity to age and metallicity makes it a suitable tracer of the old SFH \citep[e.g.,][]{Rejkuba11,Savino18,Savino20}. Due to these characteristics, the HB has long been used as a qualitative SFH indicator in nearby galaxies \citep[e.g., ][]{Dacosta96, Harbeck01, Grebel04, 2017Martin}. However, a more quantitative calibration of the HB morphology as an age indicator has so far been hindered by the complex physics of this evolutionary phase. While this issue was originally referred to as the ``second parameter problem" \citep[e.g.,][]{vandenBergh67,Sandage67}, decades of studies of Galactic globular clusters have revealed that the HB morphology is likely influenced by a large number of astrophysical parameters, the most relevant of which are stellar metallicity, age, helium abundance, and the amount of mass lost during RGB evolution \citep[e.g.,][]{Catelan09,Dotter10,Gratton10,Milone14,Tailo20}. This complexity has long hampered the use of the HB as a reliable age tracer.

However, two major developments in our understanding of stellar populations have opened a promising path to calibrate the age dependence of the HB. First, we now know that globular clusters, once thought to be the archetype of simple stellar population, have very complex chemical abundance patterns \citep[commonly reffered to as the ``multiple population" phenomenon, e.g.,][and references therein]{Bastian18}. These include large helium spreads, which have a central role in the diversity of clusters' HB morphologies \citep[e.g.,][]{Piotto07,Dalessandro11,Milone18, Tailo20}. While a small fraction of these chemically peculiar stars have been found in the Milky Way (MW) field \citep[e.g.,][]{Martell11,Schiavon17}, the multiple population phenomenon appears to be predominantly a feature of globular cluster formation. To date, there is no indication that similar abundance anomalies are present in the field population of nearby dwarf galaxies \citep[e.g.,][]{Geisler07,Salaris13,Fabrizio15}. This makes interpreting the HB of nearby dwarfs potentially simpler compared to Galactic globular clusters.

Second, we have made great strides in measuring the RGB mass loss with increasing accuracy \citep[e.g., ][]{Gratton10,Salaris13,Savino19,Tailo20}. The growing evidence is that, for old stars, RGB mass loss is a predictable function of other stellar population parameters, primarily metallicity. Furthermore, once the effect of helium abundance is removed, there is remarkable agreement between the amount of mass loss inferred from nearby dwarfs \citep{Savino19} and Galactic globular clusters \citep{Tailo20}, further supporting the argument that RGB evolution does not have a strong environmental dependence.

This means that it should be possible to use the density distribution across the HB in nearby dwarfs as a tracer of age. Specifically, as the age of the stellar population increases, the HB will progressively be occupied by hotter stars. Metallicity also affects the temperatures of HB stars, with higher metallicity producing cooler HB stars. Therefore, after factoring in the effect of metallicity, the ratio of blue HB (BHB) to red HB (RHB) stars is expected to be a good proxy for the characteristic stellar population age. The combination of age sensitivity and a luminosity that is $\sim80$ times brighter than the MSTO means that the HB has the potential to provide robust ages for galaxies out to several Mpc. This is particularly powerful in light of the thousands of faint galaxies that next generation surveys are expected to uncover in the coming decade \citep[e.g.,][]{Simon19,Mutlu-Pakdil21,Qu23}.

Here, we present a step forward toward empirically using the HB morphology to estimate the characteristic ages of distant galaxies.  Specifically, we use the MSTO-based SFHs of 27 M31 satellite galaxies from the \hst\ survey of M31 satellites to estimate the correlation of the ratio of BHB to RHB stars with mean galaxy stellar age. The M31 satellite sample provides a large, diverse, and uniformly analyzed set of SFHs and high-precision photometry that can be used to calibrate the HB as an age indicator for a variety of galaxy types.  In comparison, the MW satellites have excellent uniform photometry \citep[e.g.,][]{Munoz18a,Munoz18b}, but the SFH methodologies are inhomogeneous across the MW satellite populations \citep[e.g.][]{Lee09,deBoer14,Rusakov21}, which is not suitable for this type of analysis.  Similarly, as discussed above, the MW globular clusters have excellent photometry and well-determined ages \citep[e.g.,][]{Sarajedini07,Dotter10,Vandenberg13}, but the impact of multiple populations on their HB morphologies makes them not well-suited for direct ties to entire galaxies.  As a result, our M31 sample is among the best available for this type of work.

From our M31 data, we show that we can robustly predict the mean age of \emph{old} stars in a galaxy not in our calibration sample, using only HB morphology and an estimate of the galaxy's average metallicity. The HB provides access exclusively to star formation older than $\sim6$~Gyr, as more massive (and younger) stars do not evolve through the HB phase. The mean age of such ancient stars is useful for understanding galaxy formation mechanisms in the early Universe (e.g., quenching, connection to reionization, the ages of stellar halos in more massive systems) and for deciding which objects may warrant deeper (e.g., ancient MSTO depth) follow-up observations.  There are obvious limitations to our approach such as the insensitivity to star formation younger than $\sim6$~Gyr and the low time resolution relative to full SFHs from the oMSTO.  Nevertheless, virtually all galaxies known in the local Universe are predominantly ancient (e.g., \citealt{Weisz11}; though see \citealt{Cole07,Cole14} for a few exceptions) and even HB-depth CMDs have proven observationally challenging/prohibitive in the era of \hst\ outside $\sim1-2$~Mpc. Thus, in the broader context, a robust HB age provides an important first step in a more detailed understanding of resolved galaxies out to larger volumes.

This paper is organized as follows.  In \S \ref{sec:data} we summarize the data (i.e., photometry, SFHs) we use to calibrate the relationship between a galaxy's age, metallicity, and its HB morphology.   In \S \ref{sec:method}, we present the technical details for how we relate the HB morphology to the galaxy's age and metallicity.  In \S \ref{sec:discuss} we validate our results on datasets not used in our calibration and apply our findings to estimate the ages for a handful of galaxies outside the LG.

\section{Data}
\label{sec:data}

\subsection{Photometry and sample selection}
\label{sec:photometry}
We use \hst/ACS F475W, F606W, and F814W photometric catalogs uniformly produced by the \hst\ M31 Treasury Survey of M31 Satellites (GO-15902; PI: Weisz). The photometry used in this paper comes from new observations, taken as part of GO-15902, as well as additional archival data of the M31 system (GO-13028/13739, PI: Skillman; GO-13699, PI: Martin). The photometric reduction process with \texttt{DOLPHOT} \citep{Dolphin00} and subsequent catalog construction is summarized in \citet{Savino22, Savino23} and will be fully detailed in Weisz et al. (in prep). Here, we focus on highly complete and high signal-to-noise ratio (SNR) stars around the HB, which are robustly measured regardless of the specific choices in the reduction set-up. We use this M31 sample, as opposed to the Milky Way (MW), because of its uniformity of data quality, areal coverage, and SFH and relatively small degree of MW foreground contamination relative to the angularly larger MW satellites.  Similarly, the chemical abundance peculiarities in MW globular clusters appear to influence the morphology of the HB in ways that are different from galaxy field populations, limiting the utility of globular clusters for our purposes \citep[and references therein]{Gratton11,Dalessandro11,Bastian18}.

From the survey's full sample of 36 dwarf galaxies \citep{Savino22}, we exclude M32, NGC147, NGC185, NGC205, and \A{XIX} as their large angular extent, off-center position of the \hst\ pointings, and our incomplete knowledge of population gradients in these galaxies \citep[e.g.,][]{Rose05,Ho2015,Taibi22} make it challenging to estimate representative metallicities that are matched to our fields (cf. \S~\ref{sec:Metallicity}). Moreover, due to crowding, M32 and NGC205 do not have sufficiently deep photometry to ensure reliable stellar population ages (i.e., their CMDs do not reach the oMSTO). 

We also exclude IC1613, the Pisces dwarf (LGS3) and the Pegasus dwarf irregular, as these galaxies have had recent star formation, and contamination from bright main sequence (MS) stars makes a clean HB selection more challenging for our calibration purposes. In fact, we use the Pisces and Pegasus dwarf galaxies in \S~\ref{sec:Recent Star Formation}, to validate how our methodology performs in the presence of main-sequence contamination of the HB. We also exclude \A{IX}, which suffers from significant contamination from M31 itself, making it challenging to isolate genuine member HB stars.

%The catalogs are based on approximately \textbf{XXX} orbits of \hst\ ACS imaging, in the F475W, F606W, and F814W bands; for 29 dwarf galaxies included in the survey. The images have been reduced using \texttt{DOLPHOT} \citep{Dolphin00}, a popular point spread function photometry package, using a reduction set-up based on the PHAT survey \citep{Williams14}. Further details on the photometric reduction will be presented in a dedicated publication (Weisz et al. in prep).

% Magnitudes for Andromeda satellites in all filters were corrected for dust extinction using values from \citep{Schlafly11}. Extinction corrected color and absolute magnitude values are labeled by a subscript $0$ when used.

\begin{figure*}
\centering
\includegraphics[width=\linewidth]{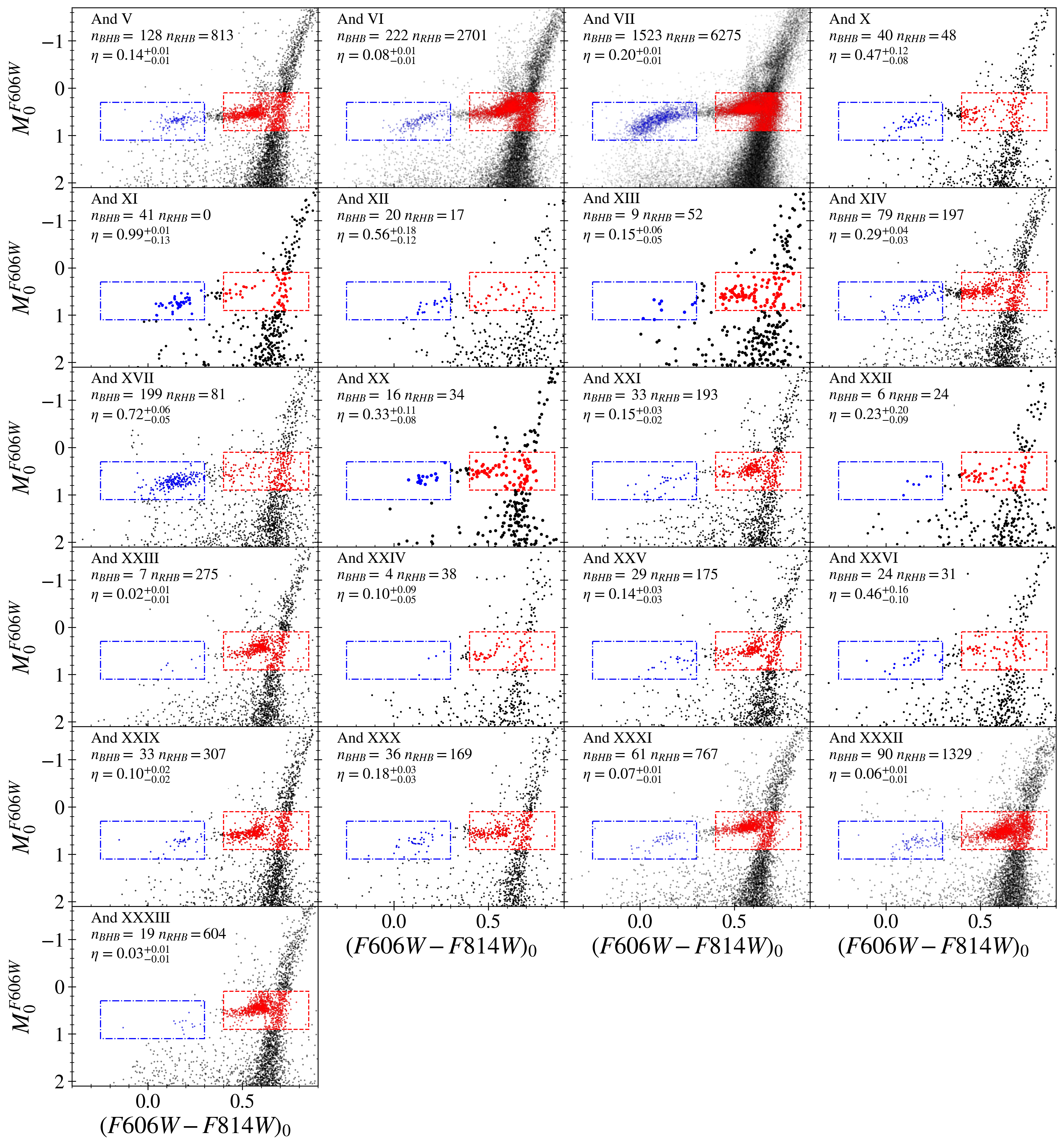}
\caption{Color-Magnitude diagrams of our M31 satellite galaxy calibrator sample with F606W/F814W data. The blue (dashed-dotted) and red (dashed) boxes show our selection regions for the BHB and RHB, respectively, as detailed in \S~\ref{sec: HB definition}. The values of $\eta=n_{BHB}/(n_{BHB}+n_{RHB})$ are also reported, where $n_{RHB}$ is found by subtracting an estimated number of RGB stars from the RHB selection box, as described in section \ref{sec:contamination}.}

% Absolute magnitudes on the y-axis assume distances from \citet{Savino22} and foreground extinctions from \citet{Green19}. Using the method described in \S~\ref{sec:contamination}, the blue box encloses the BHB (also color-coded blue) and the red box (color-coded red) captures the RHB. The value of $\eta$ is the ratio of BHB to BHB$+$RHB stars.}
\label{fig: F606 Color-Magnitude}
\end{figure*}

Figures~\ref{fig: F606 Color-Magnitude} and \ref{fig: F475 Color-Magnitude} show the CMDs of our final sample, focused on the HB region. Overall we have 21 galaxies with F606W/F814W data and 6 galaxies with F475W/F814W data.  The photometry has been transformed to the absolute magnitude space using distances from \citet{Savino22} and foreground extinction from \citet{Green19}. All the analysis in this paper will be carried out on absolute magnitudes.  The typical SNR at the HB is between 60 and 150, and the completeness, as estimated from artificial star tests, is $\sim100$\%.

\subsection{Stellar Ages}
The stellar population ages used in this work are based on a set of homogeneous SFHs measured as part of the M31 Treasury survey (Savino et al., in prep; HST-GO-15902, PI D. Weisz).  The general process of measuring these SFHs, along with the SFHs of the ultra-faint M31 satellites, are detailed in \citet{Savino23}.

In short, we derived the SFHs with the widely-used \texttt{MATCH} software package \citep{Dolphin02,Dolphin12,Dolphin13}. \texttt{MATCH} models the density distribution of stars in the CMD, down to the oMSTO in the case of all our data. We used the same distance and extinction values used in this paper \citep{Green19,Savino22}. The CMDs were modeled using the \texttt{BaSTI} stellar model library \citep{Hidalgo18}, covering an age range $10.15<\log{t}<7$, with bins of size $0.05$~dex, and a metallicity range of $-3.0<[Fe/H]<0.5$, with bins of size $0.1$~dex. The HB was explicitly excluded from the CMD fit and the age information we derive is primarily from the MSTO. The uncertainties on the SFH are then calculated using the methodology described in \citet{Dolphin12,Dolphin13}. Random uncertainties are based on Hamiltonian Monte Carlo sampling of the solution parameter space, while systematic uncertainties are estimated by introducing systematic perturbations in the shapes of the stellar models. 

For each galaxy, we use the \texttt{MATCH}-based SFH to derive the mass-weighted age of the stellar population, $t^*$, defined as:

\begin{equation}
t^*=\frac{\sum_n{t_n\cdot M_n^*}}{\sum_n M_n^*},
\end{equation}
where $t_n$ is the median age in the $n^{th}$ star formation bin and $M^*_n$ is the total stellar mass formed in that bin. As the HB is only composed of low-mass stars, we only average over the star formation bins older than 6~Gyr, which corresponds to stellar masses on the HB of roughly $1M_{\odot}$. As the precise age upper limit probed by our analysis is somewhat uncertain, we experimented with other age cuts and found that 6 Gyr provides the strongest correlation with HB morphology. The values of $t^*$ obtained in this way are listed in Tab.~\ref{table: sample}. We calculate random and systematic uncertainties on $t^*$ by sampling from the respective SFH uncertainties.

As our analysis is based on ages from a homogeneous dataset and methodology, we only use the random uncertainties in our models. This is sufficient to achieve a robust relative age calibration of the HB morphology and explore galaxy-to-galaxy differences. However, the effect of systematic uncertainties needs to be taken into account when placing our HB-based ages in a broader context. Systematic uncertainties on the value of $t^*$, inferred from the MSTO, are listed in Tab.~\ref{table: sample}. While the precise value of this systematic term changes from galaxy to galaxy, the HB-based age scale we present in this paper is derived from the full sample of galaxies. For this reason, we adopt the median value of the systematic uncertainty distribution, which is 0.86~Gyr, as our systematic uncertainty on the HB-based ages.

\subsection{Metallicities} \label{sec:Metallicity}
At present, only a modest subset of M31 satellites, typically the brightest systems, have published spectroscopic metallicities of a large number of resolved RGB stars \citep{Vargas14,Ho2015,Kirby2020}. We thus cannot rely on direct metallicity information for all galaxies in our sample.  Instead, in order to maintain uniformity across the sample, we adopt a mean metallicity value for the stellar populations in our fields using the well-established luminosity-metallicity (LZ) scaling relation from \citet{Kirby2013}:
\begin{equation}\label{eq:MetallicityKriby}
\rm \langle [Fe/H]\rangle=(-1.68\pm0.03)+(0.29\pm0.02)\log{\left( \frac{L_V}{10^6L_{\odot}}\right)},
\end{equation}
which has been obtained from spectroscopic studies of Milky Way dwarf spheroidal galaxies and that has been shown to be adequate for M31 satellites as well \citep{Ho2015,Kirby2020}. To calculate the metallicities we use stellar luminosities from \citet{Savino22}. The uncertainty on $\rm \langle [Fe/H]\rangle$ is calculated by propagating the uncertainties on the coefficients of eq.~\ref{eq:MetallicityKriby} and those on the luminosities, and adding the resulting term in quadrature with the LZ scatter of 0.17 dex reported in \citet{Kirby2013}. The values of $\rm\langle[Fe/H]\rangle$ obtained in this way are listed in Tab.~\ref{table: sample}. In Appendix~\ref{App:metallicity}, we explore the effect of adopting LZ-based metallicities, instead of spectroscopic values, and find differences compatible with the uncertainties in our model.

The value of $\rm \langle [Fe/H]\rangle$ obtained through Eq. \ref{eq:MetallicityKriby} is a mean metallicity of the galaxy stellar population \citep[as measured from RGB stars, in the original study of][]{Kirby2013}; it is not guaranteed to be representative of the HB population \citep[e.g.,][]{Nagarajan22, Savino22}, because HB and RGB stars do not come from identical stellar sub-populations and because of evolutionary timescale differences. However, direct metallicities of HB stars are impossible to obtain in any extragalactic system except the closest ones \citep[e.g.][]{Clementini05}. The next best approach is to rely on RGB-based tracers in both our calibration sample and in future applications to distant galaxies. This potential source of uncertainty is treated as a nuisance parameter in our models, as further discussed in \S~\ref{sec: MCMC fit}.   %and is captured by the intrinsic scatter term in our linear fit between age, metallicity, and HB morphology.

\begin{figure*}
\centering
\includegraphics[width=\linewidth]{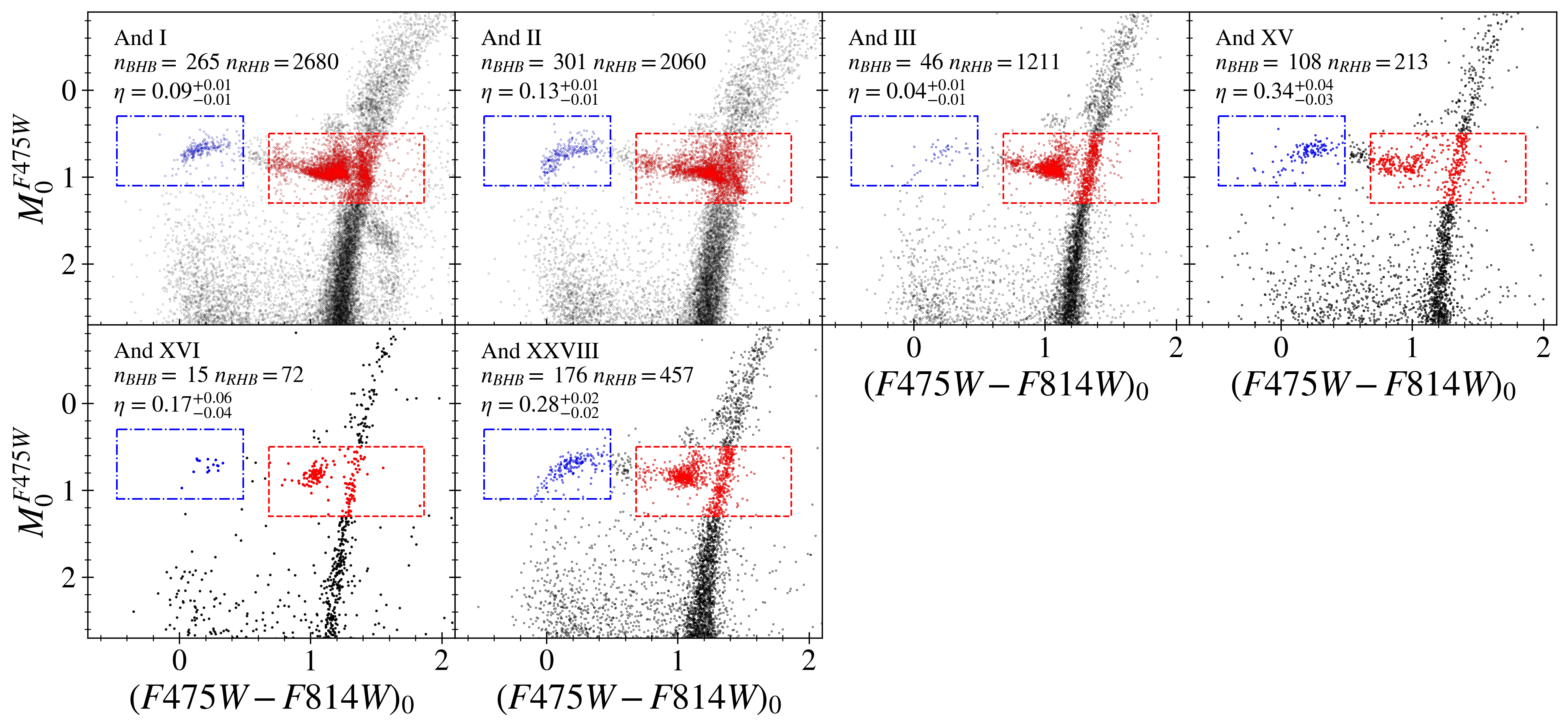}
\caption{Same as Fig.~\ref{fig: F606 Color-Magnitude} but for the galaxies with F475W/F814W data.}
\label{fig: F475 Color-Magnitude}
\end{figure*}

\begin{figure*}
\centering
\includegraphics[width=\linewidth]{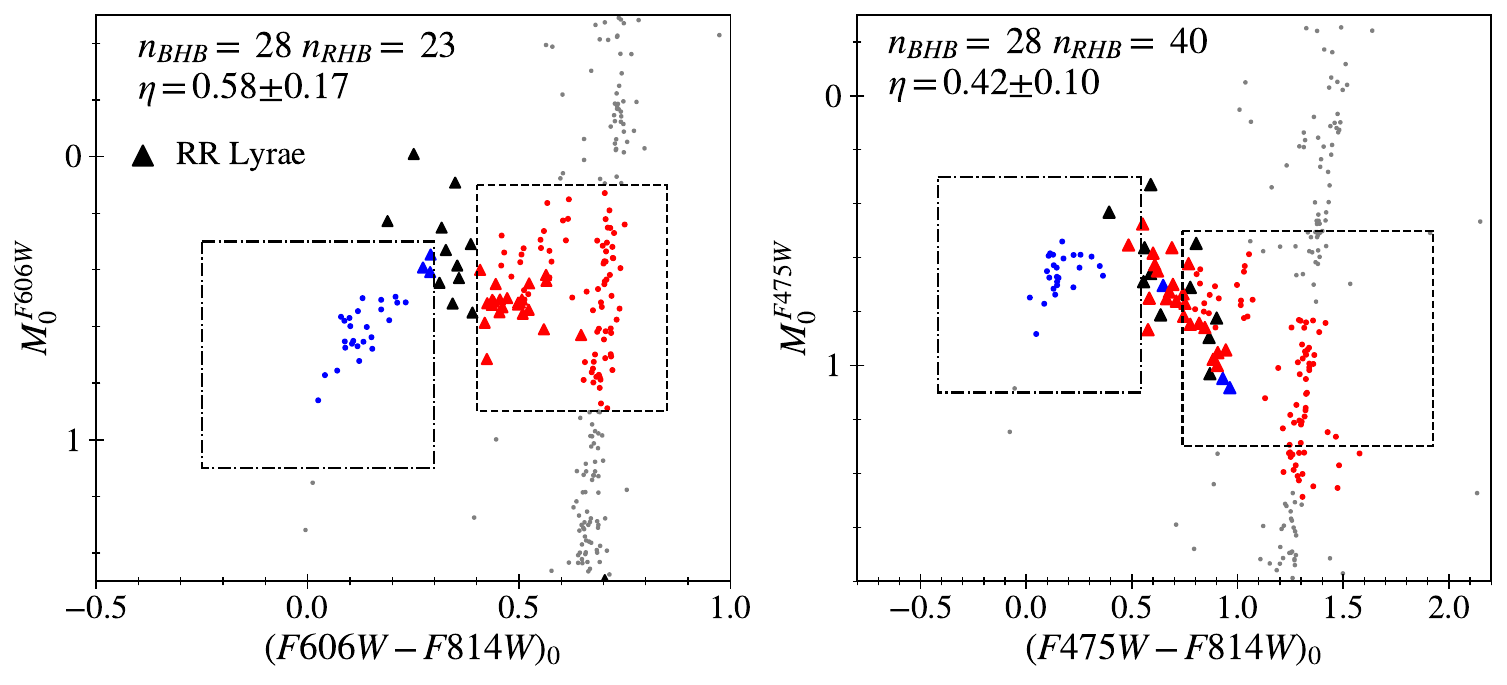}
\caption{CMD of Eridanus II zoomed in on the HB region. Left: $(F606W-F814W)_0$ vs $M_0^{F0606W}$ CMD. The dot-dashed and dashed boxes show the F606W/F814W selection regions for the BHB and RHB, respectively, as defined in \S~\ref{sec: F606W phot}. Stars inside the selection boxes are color-coded accordingly. Right: $(F475W-F814W)_0$ vs $M_0^{F475W}$ CMD. The dot-dashed and dashed boxes show the F475W/F814W selection regions for the BHB and RHB, respectively, as defined in \S~\ref{sec: F475W phot}. Stars colored blue and red are the same stars marked blue and red in the left panel using the F606W/F814W selection. RR Lyrae stars \citep{MartinezVazquez21} are marked as triangles in both plots.}
\label{fig: EriII}
\end{figure*}

\section{Methodology}
\label{sec:method}
\subsection{Measuring Horizontal Branch Morphology}\label{sec: HB definition}
The morphology of the HB can be parametrized in a number of ways \citep[e.g.,][]{Mackey05,Dotter10,Milone14}. Here we adopt a commonly used \citep[e.g,][]{Dacosta96,Dacosta00,Dacosta02,2017Martin} population ratio $\eta$, defined as:

\begin{equation}\label{eq: eta def}
\eta= n_{BHB}/(n_{BHB}+n_{RHB}),
\end{equation}
where $n_{BHB}$ and $n_{RHB}$ are the number of blue and red HB stars, respectively. The selection of BHB and RHB stars is based on the \citet{2017Martin} definition, which makes use of CMD boxes. However, we have refined our box locations to accommodate our improved photometric precision, the use of multiple photometric bands, and to explicltly address contamination from the RGB and, if present, the MS. We now provide a detailed description of our HB member selection.

\subsubsection{F606W/F814W Photometry}\label{sec: F606W phot}

The majority of our sample galaxies were observed with the F606W and F814W filter combination (Fig.~\ref{fig: F606 Color-Magnitude}). For these galaxies, we select BHB stars in the color range \mbox{$-0.25<(F606W-F814W)_0<0.3$}. This extends the BHB further to the blue from the definition used in \citet{2017Martin}. The extended BHB makes decontamination from MS stars (see \S~\ref{sec:Recent Star Formation}), if they are present, more robust. Additionally, a wide color range reduces the impact of photometric uncertainties from scattering stars outside the BHB bounds. We expect these observational uncertainties to be insignificant for galaxies in our sample, but they may become significant for more distant galaxies.
%This color range is the same used by \citet{2017Martin} and spans from the bluest BHB stars observed in our sample to the approximate blue edge of the instability strip. 

The RHB star sample spans the color range \mbox{$0.4<(F606W-F814W)_0<0.85$}. This expands the original \citet{2017Martin} definition to include the entirety of the RGB over this magnitude range. Given the variety of HB morphologies and occasional blending between the HB and the RGB, we find that this approach allows for a uniform treatment of contaminants of the RHB population, as opposed to guessing at an appropriate red edge of the selection box.  We describe this choice, and the RGB decontamination process, in \S~\ref{sec:contamination}. The vertical height of the BHB and RHB boxes is set at 0.8~mag: \mbox{$0.3<M^{F606W}_0<1.1$} for the BHB and \mbox{$0.1<M^{F606W}_0<0.9$} for the RHB. These values have been determined on our sample to roughly be centered around the HB, and are large enough to mitigate uncertainties in the distance while sufficiently small to mitigate the effect of spurious CMD contaminants. As our selection is primarily based on color, the precise magnitude range is less critical. The corresponding selection regions are shown as blue and red boxes in Fig.~\ref{fig: F606 Color-Magnitude}.

% We changed the definitions of the RHB and the BHB from the definitions used in \citep{2017Martin} to avoid contamination by stars outside the instability strip. We used AND14, which had a clear division between the RHB and the RGB as a guide in defining the edges of the RHB and BHB. In the F606W and F814W filters, the dimensions of the boxes are as follows. The magnitude of the BHB has a span of $M_{F606W} = \{0.3,1.0\}$. The color of the BHB has a span of $F606W-F814W = \{-0.1,0.3\}$. The magnitude of the RHB has a span of $M_{F606W} = \{0.1,0.8\}$. The color of the RHB has a span of $F606W-F814W = \{0.4,0.85\}$, which was chosen to extend through the Red Giant Branch. The number of contaminant stars from outside the HB is estimated using the method described in \S~\ref{sec: MS contamination}.

\begin{table*}
\centering
\caption{Values of $\eta$ measured on our calibration sample, $n_{RHB}$, $n_{BHB}$, the adopted mean metallicity, and mass-weighted age from the MSTO-based SFHs, where we ignore stars with ages less than 6 Gyrs. The HB-based mass-weighted ages from our model are also reported. Uncertianties are reported as statistical error, with the systematic error in parentheses.}%\textbf{The $n_{RHB}$ and $n_{BHB}$ reported are the mean and standard deviations of the $10^5$ Monte-Carlo realizations of $n_{RHB} = n_{red} - n_{RGB}$ and $n_{BHB} = n_{blue} - n_{MS}$ where $n_{red}, n_{blue}, n_{RGB}, n_{MS}$ are all assumed to follow poisson distributions as described in section \ref{sec:contamination}.}}

\begin{tabular}{lcccccc}

\toprule
Name & $n_{RHB}$ & $n_{BHB}$ & $\eta$ & $\rm \langle[Fe/H]\rangle$ & $t^*$ from HB  & $t^*$ from MSTO\\
&&&&dex&Gyr&Gyr\\
\toprule
And I & $2680 \pm 81$ & $265 \pm 16$ & $0.09^{+0.01}_{-0.01}$ & -1.56±0.18 & $11.12^{+0.68}_{-0.64}(\pm0.86)$ & $10.98^{+0.04}_{-0.04}(\pm0.93$) \\
And II & $2060 \pm 68$ & $301 \pm 17$ & $0.13^{+0.01}_{-0.01}$ & -1.52±0.18 & $11.17^{+0.70}_{-0.65}(\pm0.86)$ & $10.63^{+0.15}_{-0.17}(\pm0.75$) \\
And III & $1211 \pm 53$ & $46 \pm 7$ & $0.04^{+0.01}_{-0.01}$ & -1.78±0.18 & $11.23^{+0.72}_{-0.67}(\pm0.86)$ & $11.66^{+0.11}_{-0.12}(\pm0.89$) \\
And V & $813 \pm 47$ & $128 \pm 11$ & $0.14^{+0.01}_{-0.01}$ & -1.80±0.18 & $11.70^{+0.69}_{-0.66}(\pm0.86)$ & $12.40^{+0.21}_{-0.20}(\pm1.06$) \\
And VI & $2701 \pm 81$ & $222 \pm 15$ & $0.08^{+0.01}_{-0.01}$ & -1.53±0.18 & $11.02^{+0.68}_{-0.64}(\pm0.86)$ & $9.43^{+0.14}_{-0.12}(\pm1.15$) \\
And VII & $6275 \pm 127$ & $1523 \pm 39$ & $0.20^{+0.01}_{-0.01}$ & -1.34±0.19 & $10.97^{+0.76}_{-0.71}(\pm0.86)$ & $10.77^{+0.07}_{-0.08}(\pm1.16$) \\
And X & $48 \pm 16$ & $40 \pm 6$ & $0.47^{+0.12}_{-0.08}$ & -2.03±0.19 & $12.59^{+0.77}_{-0.73}(\pm0.86)$ & $12.27^{+0.17}_{-0.19}(\pm0.72$) \\
And XI & $1_{-1}^{+2}$ & $41 \pm 6$ & $0.99^{+0.01}_{-0.13}$ & -2.14±0.20 & $13.08^{+0.85}_{-0.80}(\pm0.86)$ & $12.54^{+0.24}_{-0.25}(\pm0.85$) \\
And XII & $17 \pm 8$ & $20 \pm 4$ & $0.56^{+0.18}_{-0.12}$ & -2.11±0.20 & $12.82^{+0.82}_{-0.77}(\pm0.86)$ & $12.26^{+0.27}_{-0.29}(\pm0.72$) \\
And XIII & $52 \pm 12$ & $9 \pm 3$ & $0.15^{+0.06}_{-0.05}$ & -2.09±0.19 & $12.30^{+0.77}_{-0.73}(\pm0.86)$ & $10.14^{+0.34}_{-0.36}(\pm1.15$) \\
And XIV & $197 \pm 26$ & $79 \pm 9$ & $0.29^{+0.04}_{-0.03}$ & -1.88±0.19 & $12.12^{+0.73}_{-0.69}(\pm0.86)$ & $11.35^{+0.27}_{-0.27}(\pm1.57$) \\
And XV & $213 \pm 26$ & $108 \pm 10$ & $0.34^{+0.04}_{-0.03}$ & -1.91±0.19 & $12.21^{+0.74}_{-0.70}(\pm0.86)$ & $12.11^{+0.11}_{-0.11}(\pm0.66$) \\
And XVI & $72 \pm 16$ & $15 \pm 4$ & $0.17^{+0.05}_{-0.04}$ & -2.01±0.19 & $12.19^{+0.74}_{-0.70}(\pm0.86)$ & $9.68^{+0.22}_{-0.21}(\pm0.77$) \\
And XVII & $80 \pm 22$ & $199 \pm 14$ & $0.72^{+0.06}_{-0.05}$ & -1.98±0.19 & $12.62^{+0.80}_{-0.75}(\pm0.86)$ & $12.72^{+0.12}_{-0.12}(\pm0.31$) \\
And XX & $34 \pm 10$ & $16 \pm 4$ & $0.33^{+0.11}_{-0.08}$ & -2.14±0.20 & $12.67^{+0.79}_{-0.74}(\pm0.86)$ & $12.34^{+0.32}_{-0.34}(\pm0.77$) \\
And XXI & $193 \pm 23$ & $33 \pm 6$ & $0.15^{+0.03}_{-0.02}$ & -1.85±0.18 & $11.82^{+0.70}_{-0.66}(\pm0.86)$ & $11.50^{+0.25}_{-0.25}(\pm0.92$) \\
And XXII & $24 \pm 10$ & $6 \pm 2$ & $0.23^{+0.20}_{-0.09}$ & -2.14±0.20 & $12.54^{+0.79}_{-0.75}(\pm0.86)$ & $11.81^{+0.42}_{-0.43}(\pm0.83$) \\
And XXIII & $275 \pm 28$ & $7 \pm 3$ & $0.02^{+0.01}_{-0.01}$ & -1.74±0.18 & $11.05^{+0.73}_{-0.68}(\pm0.86)$ & $10.60^{+0.36}_{-0.37}(\pm0.77$) \\
And XXIV & $38 \pm 12$ & $4 \pm 2$ & $0.10^{+0.09}_{-0.05}$ & -2.00±0.19 & $11.99^{+0.74}_{-0.70}(\pm0.86)$ & $11.68^{+0.26}_{-0.27}(\pm0.48$) \\
And XXV & $175 \pm 23$ & $29 \pm 5$ & $0.14^{+0.03}_{-0.03}$ & -1.82±0.18 & $11.76^{+0.70}_{-0.66}(\pm0.86)$ & $12.28^{+0.17}_{-0.17}(\pm0.74$) \\
And XXVI & $31 \pm 12$ & $24 \pm 5$ & $0.46^{+0.16}_{-0.10}$ & -2.18±0.21 & $12.89^{+0.83}_{-0.78}(\pm0.86)$ & $11.95^{+0.32}_{-0.44}(\pm0.77$) \\
And XXVIII & $457 \pm 38$ & $176 \pm 13$ & $0.28^{+0.02}_{-0.02}$ & -1.86±0.22 & $12.06^{+0.76}_{-0.71}(\pm0.86)$ & $10.69^{+0.18}_{-0.18}(\pm1.21$) \\
And XXIX & $307 \pm 30$ & $33 \pm 6$ & $0.10^{+0.02}_{-0.02}$ & -1.93±0.19 & $11.83^{+0.72}_{-0.68}(\pm0.86)$ & $12.38^{+0.20}_{-0.21}(\pm0.47$) \\
And XXX & $169 \pm 22$ & $36 \pm 6$ & $0.18^{+0.03}_{-0.03}$ & -1.99±0.19 & $12.15^{+0.73}_{-0.69}(\pm0.86)$ & $12.21^{+0.19}_{-0.19}(\pm0.38$) \\
And XXXI & $767 \pm 45$ & $61 \pm 8$ & $0.07^{+0.01}_{-0.01}$ & -1.53±0.20 & $11.01^{+0.69}_{-0.65}(\pm0.86)$ & $10.77^{+0.24}_{-0.28}(\pm1.15$) \\
And XXXII & $1329 \pm 54$ & $90 \pm 9$ & $0.06^{+0.01}_{-0.01}$ & -1.45±0.20 & $10.82^{+0.70}_{-0.65}(\pm0.86)$ & $11.00^{+0.13}_{-0.13}(\pm1.23$) \\
And XXXIII & $604 \pm 38$ & $19 \pm 4$ & $0.03^{+0.01}_{-0.01}$ & -1.71±0.20 & $11.05^{+0.72}_{-0.68}(\pm0.86)$ & $9.84^{+0.25}_{-0.26}(\pm0.86$) \\
\toprule

\end{tabular}

\label{table: sample}

\end{table*}

\subsubsection{F475W/F814W Photometry}\label{sec: F475W phot}
Six galaxies in our sample have data from the ISLandS program \citep{Skillman17}, which used the F475W and F814W filters (Fig.~\ref{fig: F475 Color-Magnitude}). For these galaxies, we need to ensure consistency with the sample selection that we defined for the F606W/F814W photometry, so that we can derive a homogeneous age scale. We therefore use filter transformations to map the $(F606W-F814W)_0$ color ranges defined in \S~\ref{sec: F606W phot} to $(F475W-F814W)_0$ colors. 

We derive preliminary filter transformations using theoretical zero-age HB loci from the BaSTI stellar library \citep{Hidalgo18}, which allow us to examine a large range of metallicities. Using HB tracks with $-2.2<[Fe/H]<-1.2$ , which is representative of the M31 satellite metallicity range \citep{Collins2013,Ho2015,Kirby2020}, we derive the following second-order transformation:

\begin{equation}
\begin{split}
&(F475W-F814W)_0=0.308 (F606W-F814W)_0^2 \\
&+1.735 (F606W-F814W)_0 - 0.004.
\end{split}
\label{eq:Xform1}
\end{equation}

We validate this transformation using \hst\ data for the dwarf galaxy Eridanus II, which is one of the few LG galaxies to have spatially overlapping  F475W/F814W \citep{Gallart21} and F606W/F814W \citep{Simon21} imaging. We use a photometric catalog derived with the same methodology of our primary sample \citep{Savino23} containing 497 stars, and compare the measured (F475W-F814W) color for HB and RGB stars, with the predicted values using eq.~\ref{eq:Xform1}. We find that the theoretical filter transformations are accurate, with the exception of a 0.056~mag zero-point offset. No other significant trend was found in the residuals, so we apply this term to our filter transformations, resulting in:
\begin{equation}
\begin{split}
&(F475W-F814W)_0= 0.308 (F606W-F814W)_0^2 \\
&+1.735 (F606W-F814W)_0 - 0.061.
\end{split}
\label{eq:Xform2}
\end{equation}

Through eq.~\ref{eq:Xform2}, the BHB color selection becomes \mbox{$-0.46<(F475W-F814W)_0<0.49$}, while the RHB selection becomes \mbox{$0.68<(F475W-F814W)_0<1.87$}. The vertical position of the selection boxes has been set at \mbox{$0.3<M_0^{F475W}<1.1$} for the BHB stars, and \mbox{$0.5<M_0^{F475W}<1.3$} for the RHB stars. Those values have been derived to roughly center the HB in our selection boxes.  The corresponding selection regions are shown as blue and red boxes in Fig.~\ref{fig: F475 Color-Magnitude}.

Figure~\ref{fig: EriII} shows  F606W/F814W and F475W/F814W CMDs of Eridanus II zoomed in on the HB regions. Overplotted are the HB selection regions in the two filter pairs. Our BHB selection captures virtually the same stars in both CMDs.  However, not all the RHB candidates selected in F606W/F814W are contained in the F475W/F813W RHB box, leading to a slight difference in measured $\eta$. This discrepancy is partially due to the presence of RR Lyrae stars \citep{MartinezVazquez21} in the bluest region of the RHB selection box, which experience time-dependent changes in magnitude and color. For datasets consisting of a high number of epochs, the light-curve is sufficiently sampled that the recovered magnitude and color will converge towards average values, mitigating this effect. However, for more sparsely sampled photometry, the position of RR Lyrae on the CMD can vary between datasets ovserved at different epochs.

The amount of scatter introduced by the discrete light-curve sampling depends on the number of photometric epochs, on the observing cadence, and on the photometric band, with blue bands being more affected due to the higher pulsation amplitude. For Eridanus II this is particularly a problem, as the F475W photometry only consists of 8 epochs. Using mock light curves, we estimate that such sparse sampling introduces a scatter in the F475W-F814W color of roughly 0.15 mag. This is one of the worst cases for RR Lyrae bias, but it is our only option for illustrative purposes due to a paucity of overlapping data in these two filter combinations.  Reaching the HB in most galaxies, which are more distant than Eridanus II,  requires deeper photometry and more epochs, thus making the RR Lyrae issue much less severe. For comparison, the typical effect on our M31 galaxies is of the order of 0.05 mag, for both the F606W and F475W datasets. Despite the RR Lyrae issue in Eridanus II, we note that the measured values of $\eta$ are consistent within uncertainty and that the resulting age differences (cf.\S~\ref{sec:validation}) are subdominant compared to the other uncertainty sources in our analysis as discussed below.

% Figure~\ref{fig: EriII} shows the F606W/F814W and the F475W selection boxes over the CMD of Eridanus II. While most stars fall in the same selection boxes in both filters, some stars near the edges of the selection boxes are not consistent between filters. For Eridanus II, many stars that appear to be on the RHB in the F606W filter appear in between the RHB and BHB in the F475W filter.  However, the values of $\eta$ derived from the two filter sets are consistent within $1-\sigma$, even if the mapping between the BHB and RHB stars isn't perfect.

%Eridanus II  so we can't draw conclusions about the relative errors in star counts between the RHB and BHB when changing filters.   

\subsubsection{Contamination from the RGB}\label{sec:contamination}

The RHB selection outlined in \citet{2017Martin} partially overlaps with the RGB, introducing some degree of contamination into their RHB boxes.  This is unavoidable given the modest photometric precision of their dataset. In many cases, our deeper photometry allows in many cases to define a much cleaner selection of RHB stars, avoiding contamination from the adjacent RGB (e.g., Fig.~\ref{fig: F606 Color-Magnitude}). However, instead of simply using more refined boxes, we opt to model RGB contamination for two reasons.

First, some galaxies (e.g., \A{II}, \A{VII}, \A{XXXII}) do not have a clear distinction between the reddest HB stars and the RGB, even with our high precision data. This is likely due to the higher metallicity (which shifts the HB to redder colors) and larger metallicity dispersion (which increases the color spread of both the HB and the RGB) of these galaxies.

Second, observations of galaxies outside the Local Group, for which our calibration is of particular interest, will likely not achieve sufficiently deep photometry to achieve a clean separation of the RHB and RGB. By adopting a statistical treatment of contamination from RGB stars, our method can be applied to virtually any CMD depth and HB morphology.

\begin{figure}
\centering
\includegraphics[width=\linewidth]{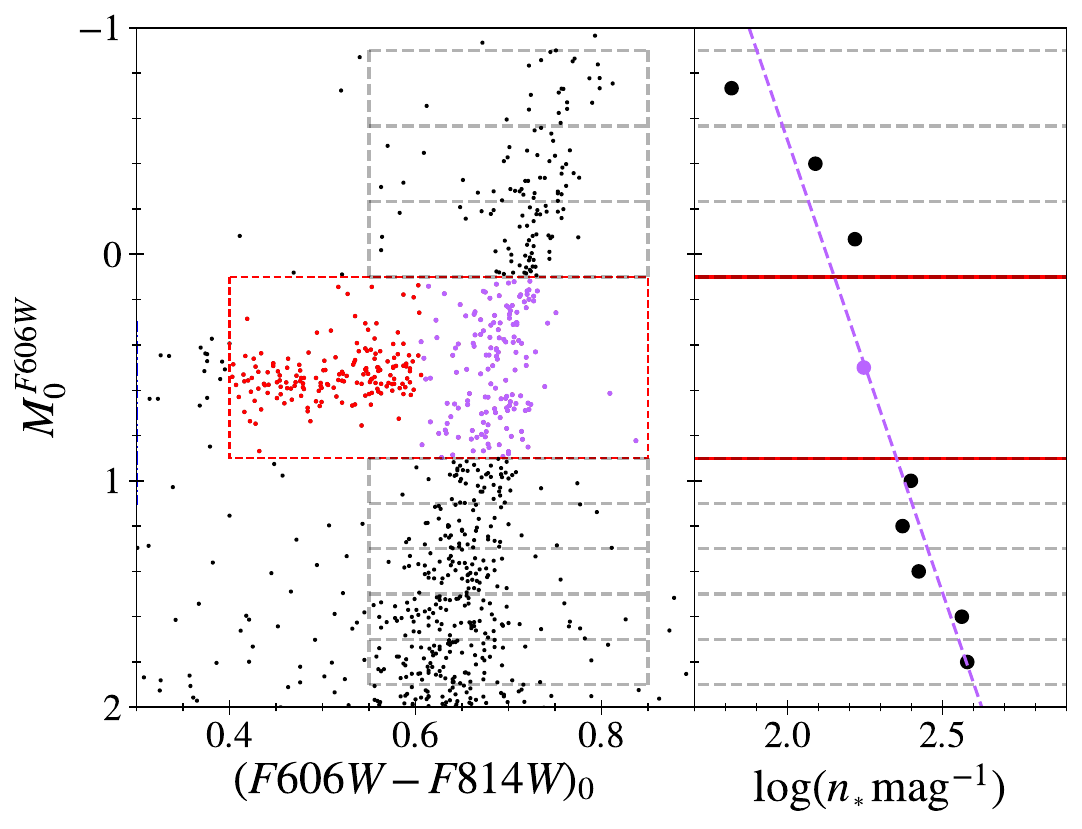}
\caption{Demonstration of our RGB decontamination process on \A{XXX}. Left: $(F606W-F814W)_0$ vs $M_0^{F606W}$ CMD of \A{XXX}. The red dashed box shows the RHB selection region, while the grey dashed regions show the selection bins to calculate the RGB luminosity function. Stars inside the RHB selection region are color-coded in order of decreasing color, with the reddest $n_{RGB}$ stars colored in purple and the remaining stars colored in red. Right: Measured RGB stellar density inside the grey dashed bins (black points). The purple dashed line shows the best fit power-law to the RGB luminosity function, while the purple point represents the actual density of RGB stars inside the RHB box, as measured from the CMD of \A{XXX}. }
\label{fig:contamination}
\end{figure}

Figure~\ref{fig:contamination} illustrates our decontamination procedure  using \A{XXX}, a galaxy that has a clear distinction between the RHB and RGB. This allows us to validate our procedure. First, we select candidate RHB stars in a box that contains both the RHB and the adjacent RGB stars. Second, we use stellar counts above and below the selection box to model the luminosity function of the RGB. The luminosity function is fit as a power law to the measured density of RGB stars in vertical bins.  Bins were defined to contain a roughly constant number of stars. The color extent of the RGB bins are \mbox{$0.55 < (F606W-F814W)_0 < 0.85$} and \mbox{$1.00 < (F475W-F814W)_0 < 1.87$}. The bins extend from $1$ magnitude above the RHB to $1$ magnitude below the RHB. Bins below the RHB have a 0.2~mag %$$\frac{1}{3}$ magnitude 
height, while bins above the RHB have a 0.3~mag height since the RGB density decreases at brighter magnitudes. 

We then integrate the model luminosity function to estimate the number of RGB stars present in the RHB selection box, $n_{RGB}$.  This value is subtracted from the total number of stars in the red box, $n_{red}$, to get $n_{RHB} = n_{red}-n_{RGB}$. We use this decontaminated $n_{RHB}$ in Eq.~\ref{eq: eta def} to calculate $\eta$. For galaxies with MS contamination of the BHB (see \S~\ref{sec:Recent Star Formation}), we use the exact same method to find $n_{BHB}=n_{blue}-n_{MS}$, where $n_{blue}$ is the total number of stars in the BHB selection box and $n_{MS}$ is the estimated number of MS contaminants.

%Some galaxies with MS contamination addressed in this paper, notably Pisces and KKR25, have a combination of a steep power law and low MS population that leads to very low MS star counts more than $0.5$ magnitudes above the BHB. For consistency, all MS decontamination was performed with bins extending $0.5$ magnitudes above and below the BHB, with 3 bins of width $0.166$ magnitude above the BHB and 5 bins of width $0.05$ below the BHB. An illustration of the extent of the MS bins is shown in Fig. \ref{fig: PiscesPeg}.} 

The uncertainty in $\eta$ is calculated using $10^5$ Monte-Carlo realizations of $n_{red}$, $n_{blue}$, $n_{RGB}$, and $n_{MS}$, which are assumed to follow a Poisson distribution, and calculating the resultant distribution of $\eta$ through eq. \ref{eq: eta def}. The $16th$ and $84th$ percentiles of this distribution are taken as our uncertainty in $\eta$. It should be noted that, in most cases, the distribution of $\eta$ is well approximated by a Gaussian. Only for values of $\eta$ close to 0 or 1 significant deviation from Gaussianity is appreciable.

%The uncertainty in $\eta$ is derived by standard quadrature error propagation. The uncertainty on $n_{BHB}$ is assumed to be purely driven by shot noise, while the uncertainty on $n_{RHB}$ combines a Poisson noise term for the total number of stars in the red box with the uncertainty on the RGB decontamination. The latter is calculated from the regression uncertainty on the power-law luminosity function model. The values of $\eta$ measured using this method for all galaxies in our sample are listed in Tab.~\ref{table: sample}.

\subsection{The HB morphology as an age indicator}\label{sec: MCMC fit}
\begin{figure*}
\centering
\includegraphics[width=\linewidth]{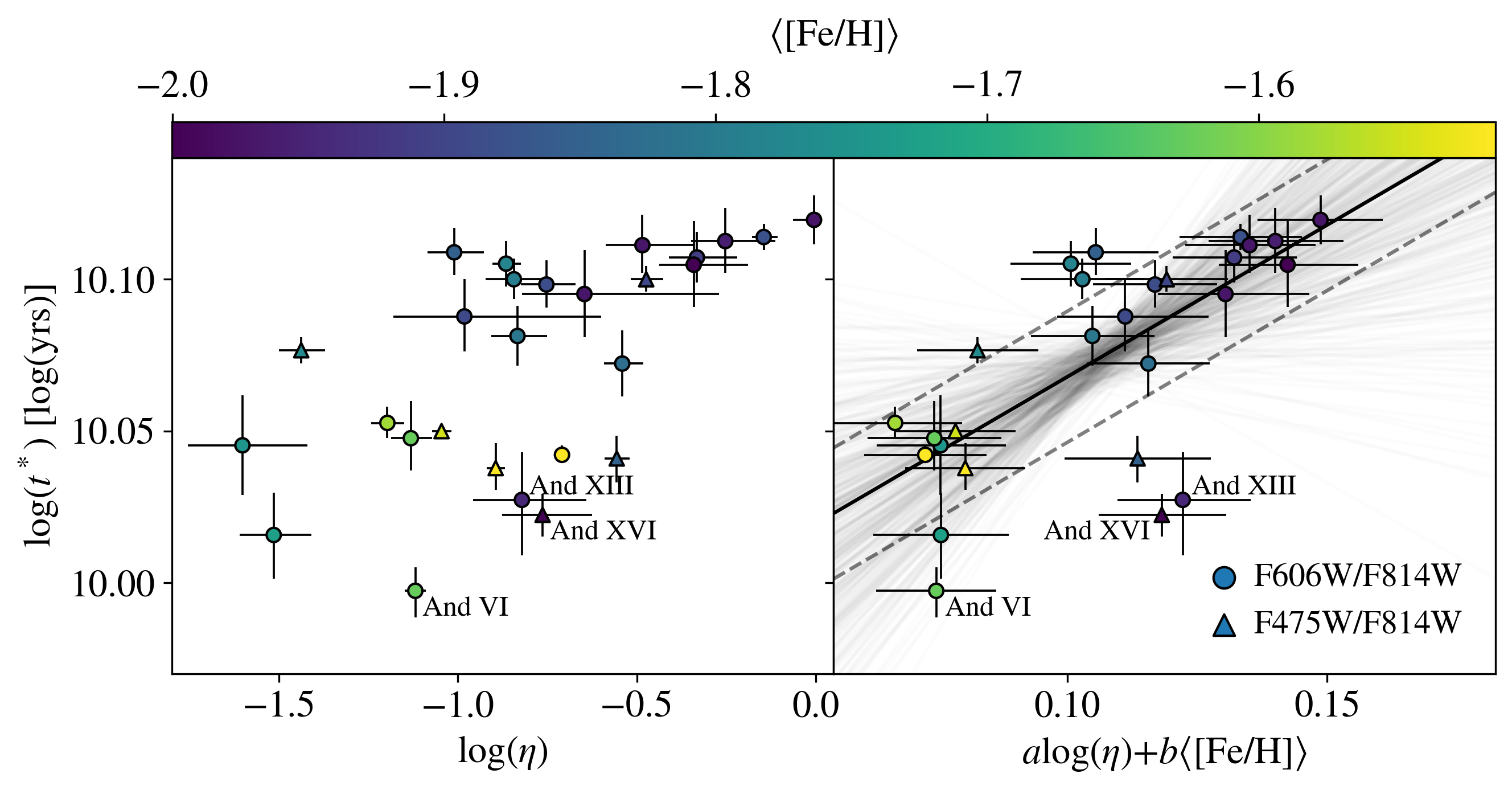}
\caption{Relationship between HB morphology ($\eta$) and mean galaxy age from the MSTO-based SFHs ($t^*$). Left:  A view of the data color-coded by metallicity. Circles and triangles indicate galaxies with F606W-F814W or F475W-F814W CMDs, respectively. Right:  Projected view of our data along the direction of minimum scatter, derived from eq.~\ref{eq:best-fit}. Overlaid are the best-fit model (solid black), draws from the posterior (light grey lines), and the width of the intrinsic scatter term in our model (dashed lines). \A{VI}, \A{XIII}, and \A{XVI} are labeled as significant outliers.}
\label{fig: MCMC line fit}
\end{figure*}

% \begin{figure}
% \centering
% \includegraphics[width=\linewidth]{Figures/eta_fe.png}
% \caption{$\eta$ and metalicity values of our calibration sample of galaxies.}
% \label{fig: eta metalicity}
% \end{figure}

The left panel of Fig.~\ref{fig: MCMC line fit} shows the mass-weighted age, $t^*$, of the old ($t>6$~Gyr) stellar population in our galaxy sample, as a function of $\eta$ and $\rm\langle[Fe/H]\rangle$ value. Consistent with expectations from stellar models and previous observational work \citep[e.g., ][]{Sarajedini95, Dotter10, Salaris13}, we observe a general trend that galaxies with a red HB morphology (low $\eta$ values) tend to have younger stellar populations, while galaxies with blue HB morphologies (high $\eta$ values) are on average older. There is also indication that, at fixed $\eta$, galaxies of higher metallicity might be comparatively younger although the trend is less defined.

A clear takeaway from Fig.~\ref{fig: MCMC line fit} is that galaxies with particularly blue HB morphologies ($\log(\eta) > -0.5$), all have predominantly old and metal-poor stellar populations, having $t^*>12$~Gyr and $\rm\langle[Fe/H]\rangle<-1.9$. This is in accordance with expectations from stellar evolution theory \citep[e.g., ][]{Cassisi13}, as hot temperatures on the HB can only be reached by low-mass metal-poor models. As the HB morphology becomes redder, the scatter of the relation increases. This is the manifestation of the dependence of the HB temperature on age and metallicity (both mean and dispersion). Similar HB morphologies, in fact, can be obtained from different combinations of age and chemical composition. 

Given this degeneracy, we model the relationship between $\eta$ and age (Fig.~\ref{fig: MCMC line fit}) using a bivariate function:
\begin{equation}\label{eq:MCMC fit}
log(t^*)  = a\cdot log(\eta) + b\cdot \langle [Fe/H]\rangle + c + \mathcal{N}(0,V),
\end{equation}
where $log(t^*)$, $log(\eta)$ and $\rm \langle[Fe/H]\rangle$ are assumed to follow a linear relation with an intrinsic scatter of variance $V$. We fit this model using the affine invariant ensemble Markov chain Monte Carlo (MCMC) sampler \texttt{emcee} \citep{Foreman_Mackey_2013_MCMC}. The free parameters of this model are $a$, $b$, $c$, as well as $ln(V)$. We adopt uniform priors listed in Tab.~\ref{table: MCMC}.

\begin{figure*}
\centering
\includegraphics[width=\linewidth]{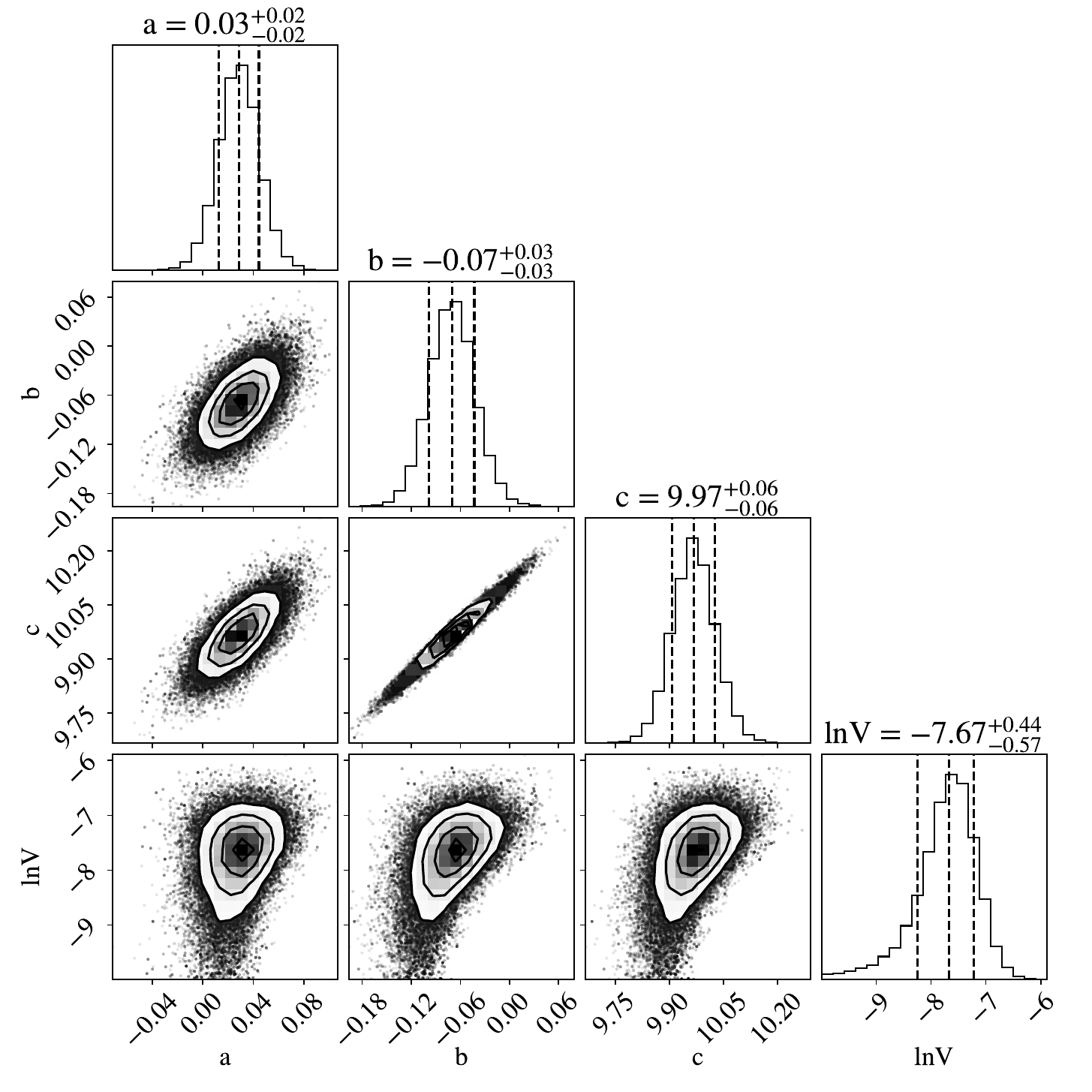}
\caption{Corner plot showing joint and marginalized posterior distributions of the parameters of Eq.~\ref{eq:MCMC fit}. The dashed lines in 1D plots mark the $15.9$, $50$, and $84.1$ percentiles of the $a$, $b$, $c$, and $ln(V)$ parameters in the posterior.}
\label{fig: MCMC eta}
\end{figure*}

We sample the posterior distribution using 32 walkers, setting the burn-in to first 100 steps, and defining the convergence length of the MCMC chain as 50 times the autocorrelation length. As both $log(\eta)$ and $log(t^*)$ have asymmetric uncertainties, we calculate our model likelihood under the assumption that these quantities follow a split-normal distribution. Figure~\ref{fig: MCMC eta} shows the corner plot of our MCMC runs. Our posterior distributions are well-constrained and unimodal. Adopting the 50th percentiles as our fiducial parameters, and the 16th and 84th percentiles as our confidence interval, the fiducial relation is:
\begin{equation}
\begin{split}
&log(t^*)  = 0.029_{-0.016}^{+0.016}\cdot log(\eta) \\ 
&-0.070_{-0.028}^{+0.027}\cdot \langle [Fe/H]\rangle + 9.967_{-0.059}^{+0.059}
\end{split}
\label{eq:best-fit}
\end{equation}
with intrinsic scatter $ln(V)=-7.671_{-0.572}^{+0.447}$, which corresponds to an \textit{rms} of $0.022$ in $log(t^*)$. For a population of 11 Gyr, this corresponds to a $0.55$ Gyr scatter. Our fiducial model is shown as a black line in the right panel of Fig.~\ref{fig: MCMC line fit}. Using eq.~\ref{eq:best-fit}, the value of $log(t^*)$ can be determined from the value of $\eta$ and $\rm \langle[Fe/H]\rangle$, with uncertainty:
\begin{equation} \label{eq: t* error}
\sigma^2_{log(t^*)}=\mathbf{J} \cdot \mathbf{C} \cdot \mathbf{J}^T + V,
\end{equation}
where $\mathbf{C}$ is the covariance matrix 
\begin{equation}\label{eq: cov matrix}
\mathbf{C} = \begin{bmatrix} 
2.66 \times 10^{-4} &   2.75 \times 10^{-4} &   7.12 \times 10^{-4} &   \\
2.75 \times 10^{-4} &   8.00 \times 10^{-4} &   1.68 \times 10^{-3} &   \\
7.12 \times 10^{-4} &   1.68 \times 10^{-3} &   3.65 \times 10^{-3} &

\end{bmatrix}
\end{equation}
which is derived from our MCMC chains, and $\mathbf{J}$ is the Jacobian matrix of Eq.~\ref{eq:best-fit}.

We comment on a few aspects of Eq.~\ref{eq:best-fit}. The first regards the equation coefficients. The coefficient of the $\eta$ term is positive, meaning that bluer HB morphologies correspond to older stellar population ages. This is evident from Fig.~\ref{fig: MCMC line fit} and is expected from stellar evolution. The coefficient of the metallicity term, on the other hand, is negative, meaning that, at higher metallicities, a given HB morphology traces a younger stellar population age. This seems at odds with stellar evolution expectations, as higher metallicities should require smaller masses (i.e., older ages) to achieve a given temperature on the HB. The interpretation of the $\rm\langle[Fe/H]\rangle$ term is more nuanced for a few reasons.

First, the metallicity obtained through Eq.~\ref{eq:MetallicityKriby} represents the stellar metallicity averaged across all stellar populations capable of producing an RGB. This is not guaranteed to be representative of the metallicity on the HB (which, given our selection, is produced only by stars older than $\sim 6$ Gyr), especially in galaxies with extended SFHs. In fact, the value of $\rm\langle[Fe/H]\rangle$, as obtained from Eq.~\ref{eq:MetallicityKriby}, is essentially a proxy for stellar luminosity, rather than a direct probe of the metallicity of the old stars.

\begin{table}
\caption{Priors and posteriors for parameters used in the model of eq.~\ref{eq:MCMC fit}.}
\centering
\begin{tabular}{cccc}

\toprule
Name & Description & Prior & Posterior\\
\toprule
a & $log(\eta)$ slope & $\mathcal{U}(-0.2,0.2)$ & $0.029_{-0.016}^{+0.016}$\\ 
b & $\rm \langle[Fe/H]\rangle$ slope & $\mathcal{U}(-0.2,0.2)$ & $-0.070_{-0.028}^{+0.027}$\\ 
c & zero point & $\mathcal{U}(9.5 ,10.5 )$  & $9.967_{-0.059}^{+0.059}$\\ 
ln(V) & intrinsic scatter & $\mathcal{U}(-10,0)$& $-7.672_{-0.572}^{+0.447}$\\

\toprule

\end{tabular}
\label{table: MCMC}

\end{table}

Second, and more importantly, ages and metallicities of stars in a galaxy are not truly independent variables. On the contrary, well defined age-metallicity relations (AMRs) are known to exist in nearby galaxies, as a result of chemical self-enrichment \citep[e.g.,][]{deBoer12a,deBoer12b}. These correlations exist within individual galaxies but also across the galaxy sample, with higher metallicity (i.e., higher luminosity) systems having on average more extended star formation histories and, therefore, younger mean ages. This is observed, for instance, in satellites of the MW \citep[e.g.,][]{Weisz15} and holds true for our M31 sample (Savino et al., in prep.). The negative coefficient of eq.~\ref{eq:best-fit} is likely capturing this correlation between galaxy mean age and metallicity. At fixed metallicity, the morphology of the HB then allows to quantify differences in mean stellar population age.

As this correlation between mean age and metallicity is encoded in our model, it follows that our calibration works best on galaxies that have broadly compatible AMRs to those of the M31 satellites. This could be a possible source of additional scatter when comparing galaxies across very different hosts. However, from eq.~\ref{eq:best-fit}, this effect only becomes larger than the intrinsic scatter term if the galaxy AMR is offset by more than approximately 0.3~dex from those of our calibration sample.

% In the ideal limit where all the galaxies in our sample shared the same AMR, the age distribution of stars in each galaxy would also be sufficient to determine its metallicity distribution, and therefore $t^*$ would be the only necessary variable to determine the HB morphology. This is clearly not the case and the $\rm\langle[Fe/H]\rangle$ (i.e., galaxy luminosity) term can be thought as encoding typical differences in the AMR, as function of luminosity, across the galaxy sample. \textbf{This further complicates the interpretation of the metallicity coefficient in terms of expectation from evolution models of single stars.} 

% For instance, in a scenario where more luminous dwarfs had slower AMRs (as a result, e.g., of more extended SFHs) compared to fainter galaxies, old stars of a given age would be more-metal poor in the former than in the latter, in spite of residing in a galaxy with higher $\rm\langle[Fe/H]\rangle$. Such a scenario would naturally result in a negative $\rm\langle[Fe/H]\rangle$ coefficient.

\begin{figure*}
\centering
\includegraphics[width=\linewidth]{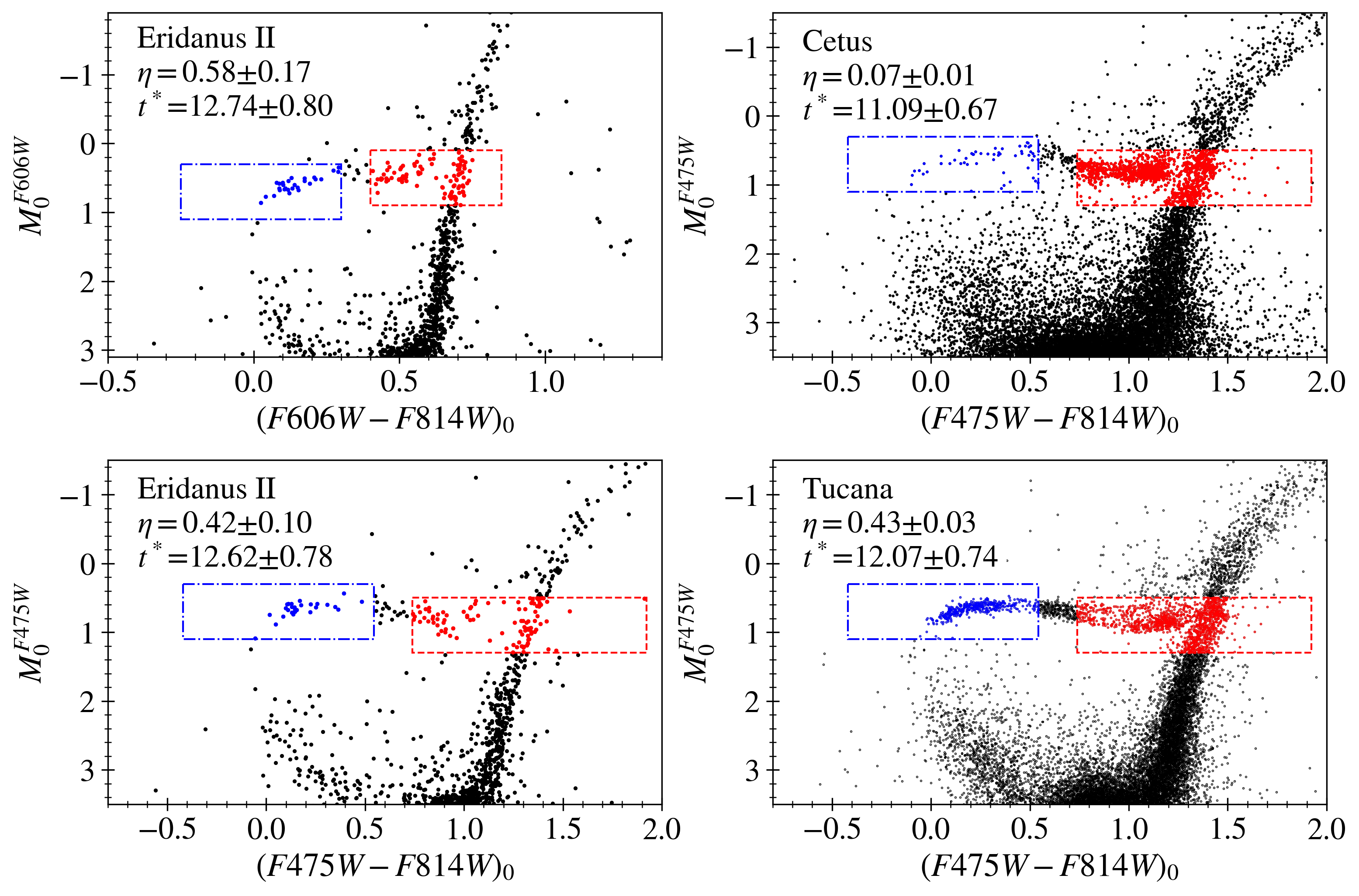}
\caption{The CMDs and BHB/RHB selection regions for our validation galaxies, Eridanus II (left panels), Cetus (upper right), and Tucana (lower right). In all cases, our inferred HB-based ages are in good agreement with the ages from the MSTO-based SFHs.}
\label{fig: LG eta}
\end{figure*}

\begin{figure}
\centering
\includegraphics[width=\linewidth]{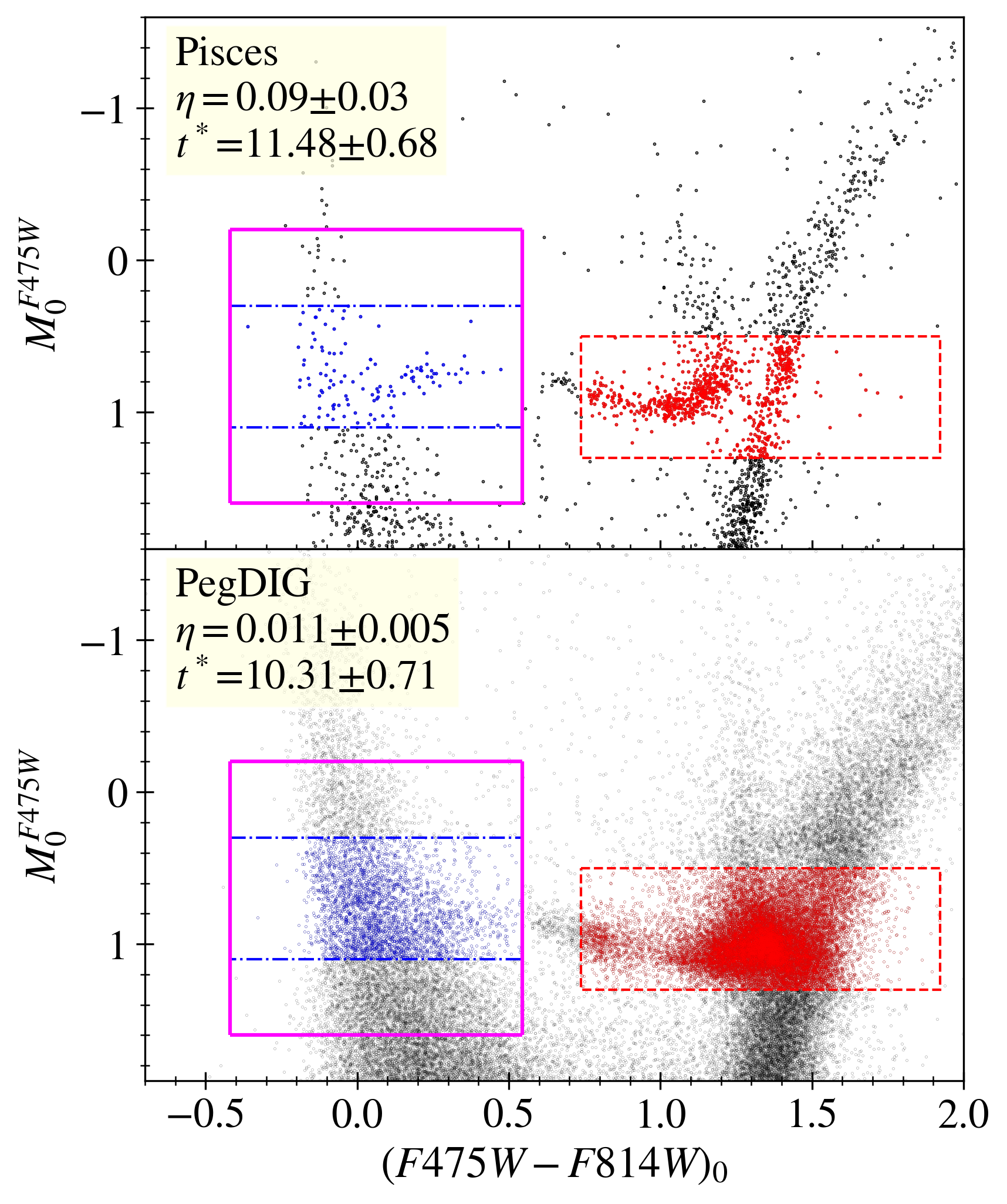}
\caption{The CMDs for the Pisces and Pegasus dwarf irregular galaxies. Main sequence subtraction, described in \S~\ref{sec:contamination}, was performed using the magenta boxes to bound the MS. Highlighted are also the BHB/RHB selection regions (blue dotted-dashed and red dashed boxes, respectively), and the regions used to remove quantify MS contamination (solid magenta boxes).}
\label{fig: PiscesPeg}
\end{figure}

Another interesting result in Eq.~\ref{eq:best-fit} regards the size of the intrinsic scatter term. The presence of a non-zero intrinsic scatter in our relation is not surprising, and this is likely arising from several sources of uncertainties in our methodology. These include the previously discussed discrepancy that could exist between $\rm\langle[Fe/H]\rangle$ and the HB metallicity. Another source of scatter may be that all the variables of our model ($t^*$, $\eta$, and $\rm\langle[Fe/H]\rangle$) describe global properties of the stellar population (being integrals of the age, HB color, and metallicity distributions, respectively). While stellar evolution suggests that a strong correlation should exist between age, metallicity, and HB color on a star-by-star basis, the correlation between global stellar population measurements might be somewhat reduced. For instance, two galaxies may have distinct stellar age distributions that result in similar values of $t^*$, while their HB morphologies might not necessarily lead to comparable values of $\eta$. Finally, we are assuming that the other variables affecting the HB (e.g., helium abundance or RGB mass loss) do not vary significantly across the galaxy sample, whereas some variation might be present.

Regardless, all these effects only introduce modest scatter in our fit. As shown in Fig.~\ref{fig: MCMC line fit}, the inclusion of a 0.022 dex $rms$ term in our model (0.5 to 0.7 Gyr, depending on the $t^*$) is sufficient to capture these additional uncertainties for virtually all galaxies (with a few exceptions discussed below). As demonstrated in \S~\ref{sec:Depth} and \S~\ref{sec: HB pop}, this intrinsic scatter is our dominant source of uncertainty on a large range of galaxy luminosities and photometric depths. The resulting age uncertainty is comparable in size to the typical systematics that affect old stellar population ages (e.g., Tab.~\ref{table: LG galaxies}), meaning that our HB calibration is robust enough to be used as an effective archaeological tracer.

\subsection{The Outliers: \A{VI}, \A{XIII}, and \A{XVI}}
The right panel of Fig.~\ref{fig: MCMC line fit} demonstrates that our model is able to capture the relation between age, metallicity, and HB morphology, for most galaxies in our sample, within measurement uncertainties and the intrinsic scatter term. However, galaxies \A{VI}, \A{XIII}, and \A{XVI} are clearly outliers relative to our model and most of the other systems. These three galaxies appear to be significantly younger than what we predict based on their HB morphology and metallicity.

Interestingly, these three galaxies also have unusual SFHs compared to the rest of the sample. As demonstrated in \citet{Savino23}, \A{XIII} has experienced unusually low levels of early star formation, compared to other M31 satellites of similar luminosity, and ignited prominent star formation much later, at $z\lesssim 3$. A similarly delayed SFH has been shown for \A{XVI} \citep{Weisz14c, Monelli16, Skillman17} and, from a recently measured SFH (Savino et al., in prep.), \A{VI} also shows a strong burst of star formation at $z\lesssim 3$.

In the context where our model described by eq.~\ref{eq:best-fit} is capturing the correlation between $t^*$ and $\langle[Fe/H]\rangle$ introduced by the AMR, it is plausible that the late dominant star formation episodes for these three galaxies resulted in a similarly unusual enrichment history. For instance, the ignition of late prominent star formation could have been fueled by the infall of fresh metal-poor gas. This would have altered drastically the AMR, with new stars forming at lower metallicity and resulting in a comparatively blue HB morphology.

At present, it is not clear how to identify whether more distant galaxies might exhibit similar behavior, save perhaps from spectroscopic data. However, we note that the incidence in our sample is low, being approximately 10\%. Under the assumption that our galaxy sample is broadly representative, which is implicit in the use of our calibration in the first place, this 10\% incidence can be considered as a rough estimate of the abundance of such objects.

\section{Discussion}
\label{sec:discuss}
\subsection{Validation on Local Group Galaxies}
\label{sec:validation}
Eq.~\ref{eq:best-fit} allows us to estimate a mass-weighted stellar population age for distant resolved galaxies for which the oldest MSTO is not accessible.  The two requirements for using this relation are (i) a CMD that reaches the HB with SNR $> 12$ (see \S~\ref{sec:Depth}) and (ii) an estimate of the mean metallicity (e.g., through the LZ relation). 

In this section, we validate our results by comparing our HB-based $t^*$ values to MSTO-based $t^*$ values for Local Group galaxies that are not in our calibration sample. Specifically, we use photometric catalogs obtained from deep \hst\ imaging of the Cetus and Tucana dSph galaxies (HST-GO-10505, PI: Gallart) and of the Eridanus II dwarf (HST-GO-14224, PI: Gallart; HST-GO-14234, PI: Simon; cfr. \S~\ref{sec: F475W phot}). The catalogs have been obtained following the same photometric reduction of our primary sample \citep{Savino23}. For a consistent comparison, we re-derived the MSTO-based SFHs using the same methodology used for our calibration sample \citep[Savino et al., in prep.]{Savino23}. This choice minimizes systematics due to, e.g., stellar models and and ensures a consistent test of our HB-based calibration. The MSTO-based SFH of Eridanus II is derived from the F606/F814W dataset.

For consistency with the calibration sample, and with application on distant galaxies, we adopted a mean metallicity value from the LZ relation of eq.~\ref{eq:MetallicityKriby} \citep{Kirby2013}. For Cetus and Tucana, we used absolute luminosities from \citet{McConnachie12}, while for Eridanus~II we used the measurement from \citet{Crnojevic16}. Within the respective uncertanties, the metallicities obtained from eq.~\ref{eq:MetallicityKriby} are in good agreement with more direct metallicity determinations \citep[e.g., ][]{Li17,Taibi18,Taibi20,Fu22}. The $M_V$ values and the derived $\rm \langle[Fe/H]\rangle$ are reported in Tab.~\ref{table: LG galaxies}. 

Figure~\ref{fig: LG eta} shows the CMDs of our validation sample and the BHB/RHB selection used to calculate $\eta$. The values of $t^*$ inferred from the HB, and those calculated from the MSTO-based SFH are reported in Tab.~\ref{table: LG galaxies}. The agreement between HB-based and MSTO-ages is excellent. In all cases, the two measurements differ by less than 0.4~Gyr, a difference that is small compared to our uncertainty budget ($\sim0.8$~Gyr).  It is also comparable to other common sources of uncertainties in the ages of resolved galaxies \citep[e.g., stellar models, distance and reddening uncertainties,][]{Gallart05}. The case of Eridanus II also demonstrates consistency between the HB-based ages from  F475W/F814W and F606W/F814W data. Despite differences in the exact BHB/RHB selection, highlighted in \S~\ref{sec: F475W phot}, the ages obtained from the two datasets differ by only 0.1~Gyr. 

\subsection{Validation on Galaxies with Recent Star Formation}
\label{sec:Recent Star Formation}
All galaxies analyzed so far in this paper have not experienced significant star-formation activity in the last few Gyr. This means that the number of sources that could contaminate our BHB selection is low. In galaxies with more recent star formation, however, massive stars on the MS will overlap in magnitude and color with the stars on the blue end of the HB, complicating the measurement of the HB morphology. In this section we apply our methodology to two such galaxies, to quantify the impact of this effect.

Figure~\ref{fig: PiscesPeg} shows the F475W/F814W CMD of the Pisces and Pegasus dwarf irregular galaxies, which are part of the M31 satellite Treasury sample. The CMDs clearly show the presence of a plume of bright MS stars in our BHB selection box. We deal with this contamination using the same method described in \S~\ref{sec:contamination}. We determine the luminosity function of the MS using stellar counts above and below the BHB box (magenta boxes in Fig.~\ref{fig: PiscesPeg}). The MS boxes span 0.5 mag in magnitude and are subdivided in five and three bins for the lower and upper boxes, respectively. The number of MS contaminants, $n_{MS}$, is then estimated by fitting the luminosity function with a power-law and integrating the resulting luminosity density over the span of the BHB box. The decontaminated sample of BHB stars is then used to calculate $\eta$.

The values of $t^*$ we measure through the HB morphology of Pisces and Pegasus are $11.48^{+0.70}_{-0.66}$~ Gyr and $10.31^{+0.74}_{-0.69}$~Gyr, respectively (assuming absolute luminosities from \citealt{Savino22}). We compare these values with the corresponding $t^*$ obtained from the MSTO (Savino et al., in prep., listed in Tab.~\ref{table: LG galaxies}) and find that the two measurements differ by approximately 0.6~Gyr. This is somewhat larger than the difference measured for passive galaxies in \S~\ref{sec:validation}, which reflects the added uncertainty introduced by the MS decontamination. Nevertheless, HB and MSTO based ages are still compatible to the level of both statistical and systematic uncertainties. The conclusion is that, while the presence of the bright MS might have an effect on the fidelity of HB-based ages, the accuracy of our method is still marginally impacted.

% We tested our fit (Eq. \ref{eq:MCMC fit}) on Eridanus II dwarf, Cetus, and Tucana. Eridanus II has data in both the F475W and F606W filters, so values of $\eta$ and $t^*$ were calculated using both. We also tested our results on KKR25 and VV124, which are nearby galaxies that fall outside the local group. Both of these galaxies had significant main sequence contamination, which we removed using the method described in \S~\ref{sec: MS contamination}. 

% For Eridanus, Cetus, and Tucana, the values of $t^*$ found using the Horizontal Branch are in agreement with the values from the Main Sequence Turn-off. KKR25 and VV124 have no estimated MSTO ages in the literature, so the values presented here are the first estimates of the ages of those galaxies.

\begin{figure}
\centering
\includegraphics[width=\linewidth]{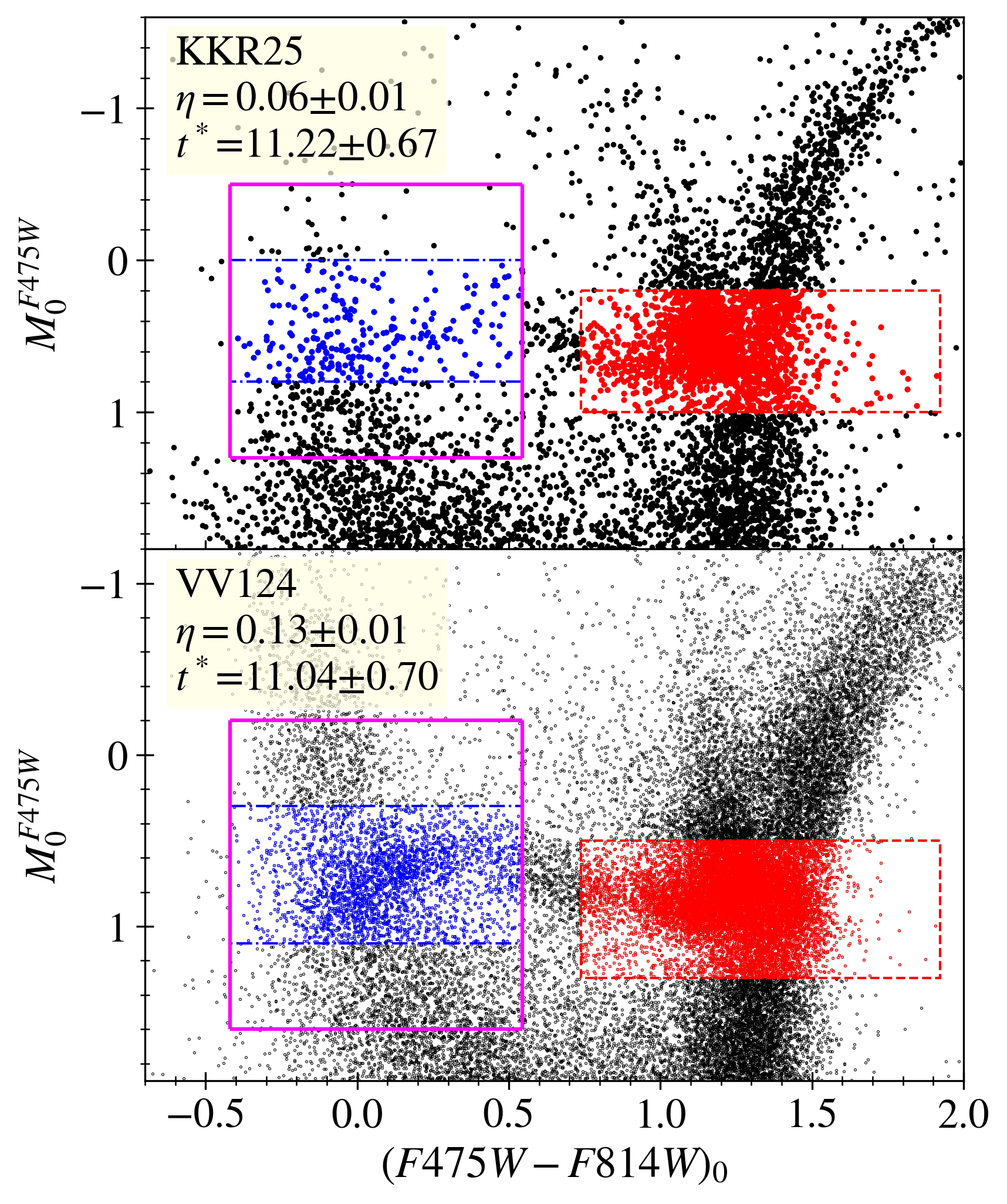}
\caption{Same as Fig.\ref{fig: PiscesPeg}, but for the dwarf galaxies KKR~25 and VV~124.}
\label{fig:KKVV}
\end{figure}

\begin{table*}
\caption{ HB-based mean ages for select nearby galaxies.  In some cases, we are able to directly compare to MSTO-based ages from CMD fitting and find excellent agreement. For the galaxies with no MSTO age listed their CMDs do not reach the MSTO.  Our age estimates are among the first in the literature for these systems.}
\centering
\begin{tabular}{lccccccc}

\toprule
Name & $(m-M)_0$ & $D$ & $\eta$ & $M_{V}$& $\rm \langle[Fe/H]\rangle$ &$t^*$ from HB & $t^*$ from MSTO\\
&mag&Mpc&&mag&dex&Gyr&Gyr\\
\toprule
EriII(F606W) & 22.67±0.04 & 0.34±0.01 & $0.58^{+0.21}_{-0.14}$ & -7.1 & -2.06±0.19 & $12.73^{+0.82}_{-0.77}(\pm0.86)$ & $12.41\pm0.24$ \\
EriII(F475W) & 22.67±0.04 & 0.34±0.01 & $0.42^{+0.12}_{-0.08}$ & -7.1 & -2.06±0.19 & $12.62^{+0.80}_{-0.75}(\pm0.86)$ & $12.41\pm0.24$ \\
Cetus & 24.46±0.12 & 0.78±0.04 & $0.07^{+0.01}_{-0.01}$ & -11.2 & -1.58±0.18 & $11.10^{+0.69}_{-0.65}(\pm0.86)$ & $10.75\pm0.08$ \\
Tucana & 24.74±0.12 & 0.89±0.05 & $0.42^{+0.02}_{-0.02}$ & -9.5 & -1.78±0.18 & $12.06^{+0.76}_{-0.72}(\pm0.86)$ & $12.06\pm0.12$ \\
Pisces & 23.91±0.05 & 0.61±0.01 & $0.09^{+0.03}_{-0.03}$ & -9.8 & -1.74±0.18 & $11.49^{+0.70}_{-0.66}(\pm0.86)$ & $10.87\pm0.19$ \\
PegDIG & 24.74±0.05 & 0.89±0.02 & $0.01^{+0.01}_{-0.01}$ & -12.3 & -1.45±0.18 & $10.31^{+0.73}_{-0.68}(\pm0.86)$ & $11.14^{+0.09}_{-0.10}$ \\
KKR25 & 26.42±0.07 & 1.92±0.06 & $0.06^{+0.01}_{-0.01}$ & -10.5 & -1.66±0.18 & $11.21^{+0.69}_{-0.65}(\pm0.86)$ & - \\
VV124 & 25.67±0.11 & 1.36±0.07 & $0.13^{+0.01}_{-0.01}$ & -12.5 & -1.43±0.18 & $11.04^{+0.73}_{-0.68}(\pm0.86)$ & - \\
\toprule

\end{tabular}
\label{table: LG galaxies}

\end{table*}

\subsection{Application to Data Beyond the Local Group: KKR~25 and VV~124}
\label{sec:KKVV}
The ability to obtain reliable stellar ages from the morphology of the HB is of particular value for galaxies in the immediate vicinity ($1 \lesssim D \lesssim 5$ Mpc) of the Local Group. Detecting the oMSTO of galaxies in this distance range is either impossible or impractically expensive with current telescopes. On the other hand, the brighter and less crowded HB stars can be observed much more efficiently in these galaxies and represent the best available tracer of their early star formation. In this section, we provide a first application of our new method by deriving stellar ages for the isolated galaxies VV~124 \citep[$D\sim 1.37\pm0.07$ Mpc,][]{Tully13} and KKR~25 \citep[$D\sim 1.93\pm0.06$ Mpc,][]{Makarov12}.

VV~124 and KKR~25 are two relatively bright ($10^6\lesssim L/L_{\odot} \lesssim 10^7 $), metal-poor ($-1.9 \lesssim \langle [Fe/H]\rangle \lesssim -1.5$) dwarf galaxies located just outside the LG \citep[e.g.,][]{Karachentsev01,Kopylov08,Bellazzini11a,Makarov12, Kirby12, Kirby13b}. Both galaxies host predominantly old/intermediate stellar populations with a small amount of recent $(\lesssim 1$Gyr) star formation \citep{Kopylov08,Bellazzini11b,Makarov12,Neeley21}, have relatively low $\rm H_{I}$ content \citep{Begum05,Bellazzini11a}, and (in the case of VV~124) pressure-supported kinematics \citep{Kirby12}.  Many of these observations have led to suggestions that these galaxies are transition dwarfs in a late stage of their evolution.

The properties of KKR~25 and VV~124 are particularly intriguing in the context of low-mass galaxy evolution. The long-established morphology-density relationship has been central to the idea that environment is a primary driver in the evolution and quenching of luminous ($L \gtrsim 10^6 L_{\odot}$) dwarf galaxies \citep[e.g., ][]{Mateo98, Grebel03, Geha12,Putman21}. Yet, early-type dwarfs such as KKR~25 and VV~124, are distant enough from other massive galaxies that past interactions are virtually impossible. This poses a challenge in our understanding of why such objects have relatively low levels of recent star formation and small neutral gas reservoirs in light of environmentally-driven theories of galaxy quenching.

While the distances of these two galaxies make their oMSTOs very challenging to observe, their HBs are well within the reach of existing \hst\ observations. To measure the HB morphology of these two targets we have therefore performed PSF-photometry (using the same procedure as in \S~\ref{sec:photometry}) on the archival \hst\ data of program GO-15244 (PI: Monelli).  Each galaxy has 16 orbits of ACS F475W and F814W imaging that have previously been used to study their RR Lyrae \citep{Neeley21}.

Figure~\ref{fig:KKVV} shows the CMDs of KKR~25 and VV~124 around the HB region. The helium burning sequences of these galaxies are extended from the blue end of the HB to more massive helium-burning stars above the red clump, suggestive of an extended SFH. A notable feature in this figure is that both galaxies have a bright plume of young main-sequence (MS) stars which overlaps with the BHB. We account for this using the same procedure described in \S~\ref{sec:Recent Star Formation}.

\begin{figure*}
\centering
\includegraphics[width=\linewidth]{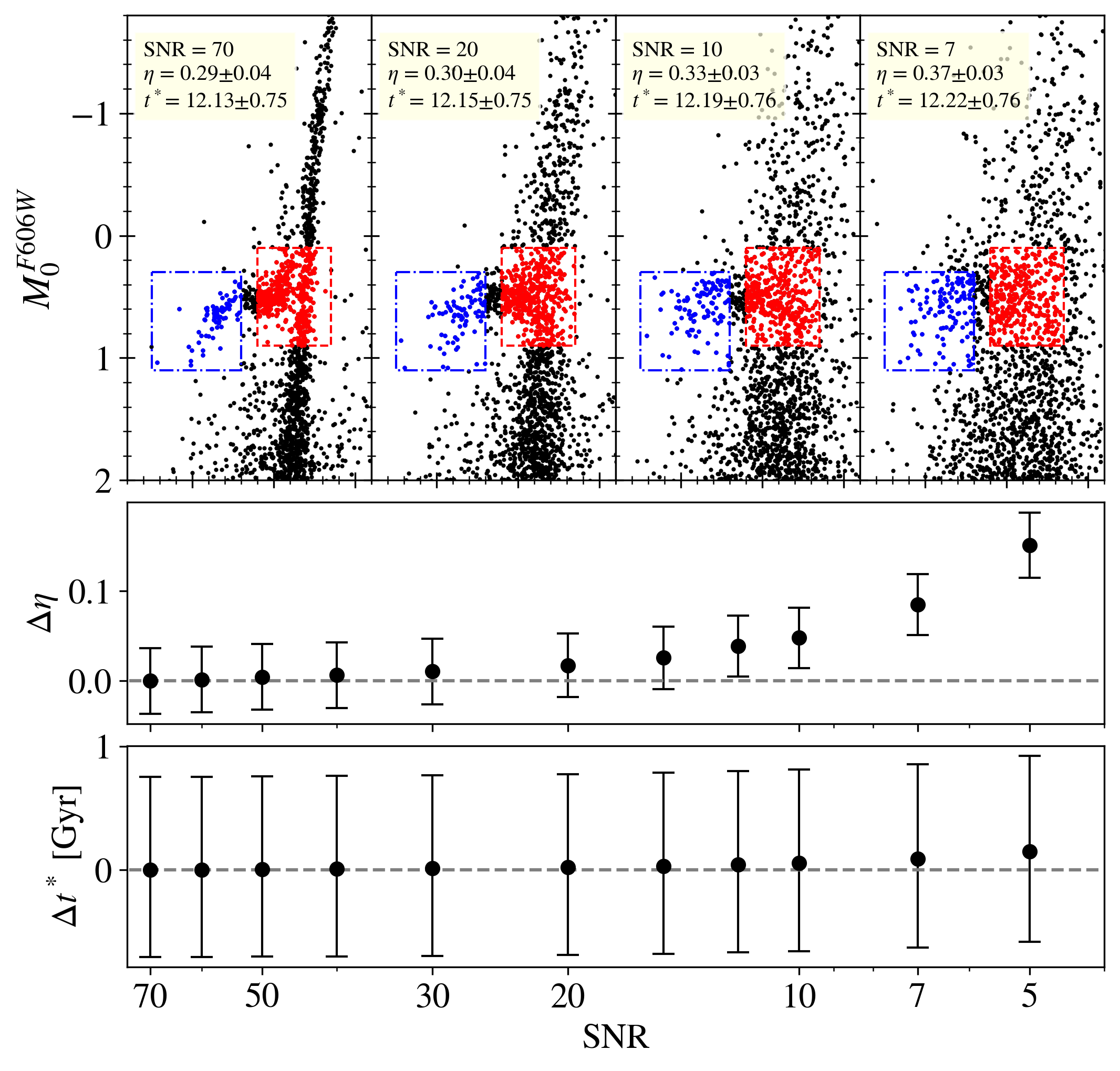}
\caption{Dependence of HB-based age accuracy on photometric depth. Upper panel: example CMDs of \A{XIV}, obtained by artificially decreasing the SNR. Selected BHB  (blue) and RHB (red) are highlighted. Middle panel: Difference, with respect to the original photometry, in the measured $\eta$ of \A{IV} as a function of SNR. Lower panel: Difference, with respect to the original photometry, in the measured $t^*$ of \A{IV} as a function of SNR. The values of $\eta$ and $t^*$ can be reliably recovered down to SNR$\gtrsim 10$ on the HB.}
\label{fig: SNR}
\end{figure*}

% To mitigate the young MS contamination, we have applied an analogous approach to that described in \S~\ref{sec:contamination} with the RGB contamination. We have derived the luminosity function of the MS using two selection boxes above and below the BHB selection box (cyan boxes in Fig.~\ref{fig:KKVV}). The MS boxes cover the full color extent of the BHB box, i.e. $-0.23<(F475W-F814W)_0<0.49$, and have a vertical height of 0.5 mag. The luminosity function of the MS is derived by binning the stellar counts in bins that are 0.1 and 0.33 mag wide, in the lower and upper box respectively. The measured stellar counts are then fit with a power law and the resulting best fit is used to estimate the number of MS contaminants in $n_{BHB}$. As in the case of the RGB contamination, the uncertainty in this correction term is added in quadrature to the shot noise term to estimate the total uncertainty in $n_{BHB}$.

The resulting HB morphology parameters are $\eta=0.063\pm0.014$ for KKR~25 and $\eta=0.132\pm0.007$ for VV~124. Using absolute magnitudes from \citet{McConnachie12} and deriving average metallicities through eq.~\ref{eq:MetallicityKriby} (listed in Tab.~\ref{table: LG galaxies}), we find $t^*=11.21^{+0.69}_{-0.65}$~Gyr for KKR~25 and $t^*=11.03^{+0.73}_{-0.68}$~Gyr for VV~124. These mass-weighted mean ages are similar to Cetus and Pisces, and quite younger than Tucana and Eri~II.  

Given that $t^*$ is an integrated measurement of the SFH, the interpretation of the mean ages of KKR~25 and VV~124 depends on our assumptions on when star formation began in these objects. Under the assumption that star formation began at the earliest epochs, our age measurements indicate a substantially extended SFH, akin to the aforementioned Pisces and Cetus dwarfs \citep{Monelli10,Hidalgo11}. On the other hand, it is possible that star formation remained low for the first one or two Gyr, and then experienced a short duration burst around $z\sim3$. While we cannot discard this option, we note that VV~124 presents a prominent BHB population (clearly seen through the MS contamination), which suggest the presence of a sizable population of truly ancient stars.

Regardless, the presence of significant star formation at $\sim 11$ Gyr suggests that reionization had little impact on the early star formation of these galaxies.  In comparison, isolated galaxies Leo~A and DDO~210 have similar present day luminosities, but only formed 10\% of their stellar mass at ages $>$10~Gyr, which, in some interpretations, could be due to suppression from the cosmic ultra-violet background \citep[e.g.,][]{Cole07, Cole14}.  Similarly, the degree of isolation of these galaxies, combined with their star formation activity at low ($\lesssim 3$) redshift, suggests that the transition from gas-rich to gas-poor dwarf galaxies cannot solely be a function of environment.  According to the 2023 version of the Nearby Galaxy Catalog \citep{Karachentsev04,Karachentsev13}, the nearest massive neighbors to KKR~25 and VV~124 are more than 1~Mpc away (UGC8508 and Sextans B, for KKR~25 and VV~124, respectively), and even those are relatively low-mass dwarfs ($M_*\sim 10^8M_{\odot}$). Any environmental interaction these galaxies might have had with a large galaxy must have happened at ancient times and, if responsible for their nature of early type galaxies, would have resulted in a very early depression of star formation. 

For reference, the Cetus dSph, which has been proposed as a candidate backsplash galaxy from the MW or M31 \citep[e.g.,][]{Sales07,Teyssier12}, has a similar $t^*$ to VV~124 and KKR~25 but it currently lies at $\sim 700$~kpc from either large spirals \citep{McConnachie12}. This is roughly two times closer than VV~124 and three times closer than KKR~25. Tucana, another candidate backsplash galaxy, lies at a larger distance of $\sim 900$~kpc from the MW \citep{McConnachie12}, but has a $t^*$ that is $\sim 1$~Gyr older. It therefore seems unlikely that VV~124 and KKR~25 could have had an interaction with the MW or M31 and managed to sustain star formation for as long as their HB seems to indicate.

Finally, if, these galaxies had ancient SFHs that are more like MW satellites, i.e., dominant ancient episodes, than other isolated galaxies (e.g., Leo~A, DDO~210), then they provide interesting counter examples to the slow/fast scenario proposed by \citet{Gallart15}, in which the dominance of ancient star formation can be mapped to the local environmental density at high-redshifts.  While this limited sample hints at new insight into isolated dwarf galaxy formation, a large systematic study (e.g., if HB-depth photometry were available for the entire ANGST sample; \citealt{Dalcanton09}) is needed to enable a truly statistical study.

\subsection{Application to Data Beyond the Local Group: The Effect of Photometric Depth}
\label{sec:Depth}
A challenge of characterizing the HB morphology of galaxies beyond the LG will be the limited photometric precision that observations of these distant targets will achieve at the HB magnitudes. In contrast, our calibration sample has exquisite photometry of the HB region, with mean SNR ratios between 147 (\A{XVI}) and 58 (\A{XXVI}). To quantify the effect of lower photometric precision on the recovered ages, we have artificially degraded the photometry of one of our calibration galaxies and re-applied our methodology.

Figure~\ref{fig: SNR} illustrates our experiment. We start with the original photometry of \A{XIV} as our ground truth. The mean magnitude SNR of this dataset, at the magnitude level of the HB, is 74. The measured HB morphology index is $\eta=0.29\pm0.04$, which results in an age of $12.12\pm0.71$ Gyr. We then perturbed the photometry of individual stars in \A{XIV} with a random Gaussian term $\mathcal{N}(0,\sigma)$. The term $\sigma$ is chosen to emulate a target SNR: $\sigma = \sqrt{\frac{1}{SNR^2}-\frac{1}{74^2}}$. We adopt the same $\sigma$ for the whole photometry, as we are considering a relatively small magnitude range in the CMD. The upper panel of Fig.~\ref{fig: SNR} illustrates select examples of the CMDs we obtain. We then measure $\eta$ from the new degraded CMDs and derive $t^*$ values using eq.~\ref{eq:best-fit}.

Because the morphology of the HB is parameterized using large selection boxes, the value of $\eta$ only begins to significantly change as a function of SNR for SNRs as low as $\sim10$. We find that below SNR$\sim10$, the measured value of $\eta$ can deviate significantly from the truth, in excess of measurement uncertainties. Though the $t^*$ is not heavily affected in our example, because the scatter term in Eq.~\ref{eq: t* error} dominates, we find that the shallowest photometric depths affect our accuracy to $\sim 200$~Myr.  Moreover, at such low SNRs, the lower RGB will no longer be accessible, meaning that our decontamination will not work.  In contrast, for HBs with SNR $\gtrsim 10$, $\eta$ and age are statistically consistent and the lower RGB can be accessed for decontamination purposes.  

%In contrast, the change in $\eta$ between our highest and lowest SNR CMDs only changes $t^*$ by 0.3~Gyr. We therefore find that our methodology can be applied to datasets with virtually any photometric precision.

% To test for the effects of photometric uncertainty, $\epsilon$, in HB stars on $t^*$ we used \A{XIV} which had low photometric uncertainty. We adjusted the magnitude of each star in both filters to be $M'=M+\mathcal{N}(0,\epsilon)$, where $\mathcal{N}(\mu,\sigma)$ is a normal distribution. $\epsilon$ was chosen for each instance to bring the total photometric uncertainty to the desired signal-to-noise ratio, $\epsilon=\sqrt{\sigma_M^2-\frac{1}{SNR^2}}$ Using the new values of $M'$, we calculated $\eta$ using eq.~\ref{eq:eta def} and $t^*$ using eq.~\ref{eq:MCMC fit}.

% For low photometric noise, the errors in $\eta$ and $t^*$ remain low. The observed error in $t^*$ due to photometric uncertainty remains less than the combined error due to Poisson noise and fit uncertainty up to a magnitude error of $~0.1$, corresponding to a signal-to-noise ratio of 10. Past this, the resulting error in $\eta$ begins to increase significantly. The average magnitude error for the galaxies used to fit eq.~\ref{eq:MCMC fit} was $0.01$, so we don't expect that photometric uncertainty is a leading source of error in that equation. 

\subsection{Application to Data Beyond the Local Group: The Effect of CMD Population}
\label{sec: HB pop}

Thanks to future deep, large-area surveys (e.g., Rubin, Roman, Euclid), our census of low-mass, nearby galaxies is poised to dramatically increase in the coming decade.  For example,  a wealth of new low-mass galaxies down to  $M_V\sim -5$ within $\sim5$~Mpc should be discoverable \citep[e.g.,][]{Mutlu-Pakdil21}. Characterizing the ages of these galaxies from the MSTO will be challenging due to the large distances.  While the HB will be much brighter, it may also be sparsely populated (e.g., a few dozen HB stars for bright ultra-faint dwarfs).  Accordingly,  it is important to quantify the accuracy of our method as a function of stellar population size. 

To do so, we start by using two of our calibration galaxies, with different values of $\eta$, for ground truth. We have selected \A{V} ($M_V=-9.3$; \citealt{Savino22}), which has a population of $n_{red}+n_{blue}=1651$ stars (before decontamination) and a relatively red HB morphology ($\eta = 0.14$), and \A{XVII} ($M_V=-7.8$; \citealt{Savino22}), which has a population of $n_{red}+n_{blue}=479$ stars and a much bluer HB morphology ($\eta=0.72$). We then re-sample (without replacement) the CMD of these two galaxies until a given number of HB stars, $n_{HB} = n_{red}+n_{blue}$, is reached. For a given $n_{HB}$ we generate 1000 CMD realizations and calculate the corresponding $\eta$ and $t^*$ values. We also scale the absolute luminosity of both galaxies accordingly, as we decrease the HB counts, to estimate the approximate galaxy luminosity at which HB sampling becomes relevant.

% \begin{figure*}
% \centering
% \includegraphics[width=\linewidth]{Figures/HB_pop_both.png}
% \caption{$\Delta \eta$ and $\Delta t^*$ of \A{V} and \A{XVII} as HB population size is decreased. Black points and error bars show the mean and standard deviation of $\eta$ and $t^*$, calculated from the 1000 Monte Carlo samplings of the HB of \A{V}. Green dashed lines (top panel) show the analytical uncertainty on $\eta$, obtained through the procedure described in section \ref{sec: HB definition}. The blue dashed lines (bottom panel) show the extent of the intrinsic scatter term, from eq~\ref{eq:MCMC fit}. The uncertainty in $t^*$ is dominated by the intrinsic scatter term for $n_{\rm BHB}+n_{\rm RHB}$ $> 30$, which corresponds to an absolute magnitude of $M_V < -5$.}
% \label{fig: HB population}
% \end{figure*}

\begin{figure*}
    \subfloat
        {\includegraphics[width=0.5\linewidth]{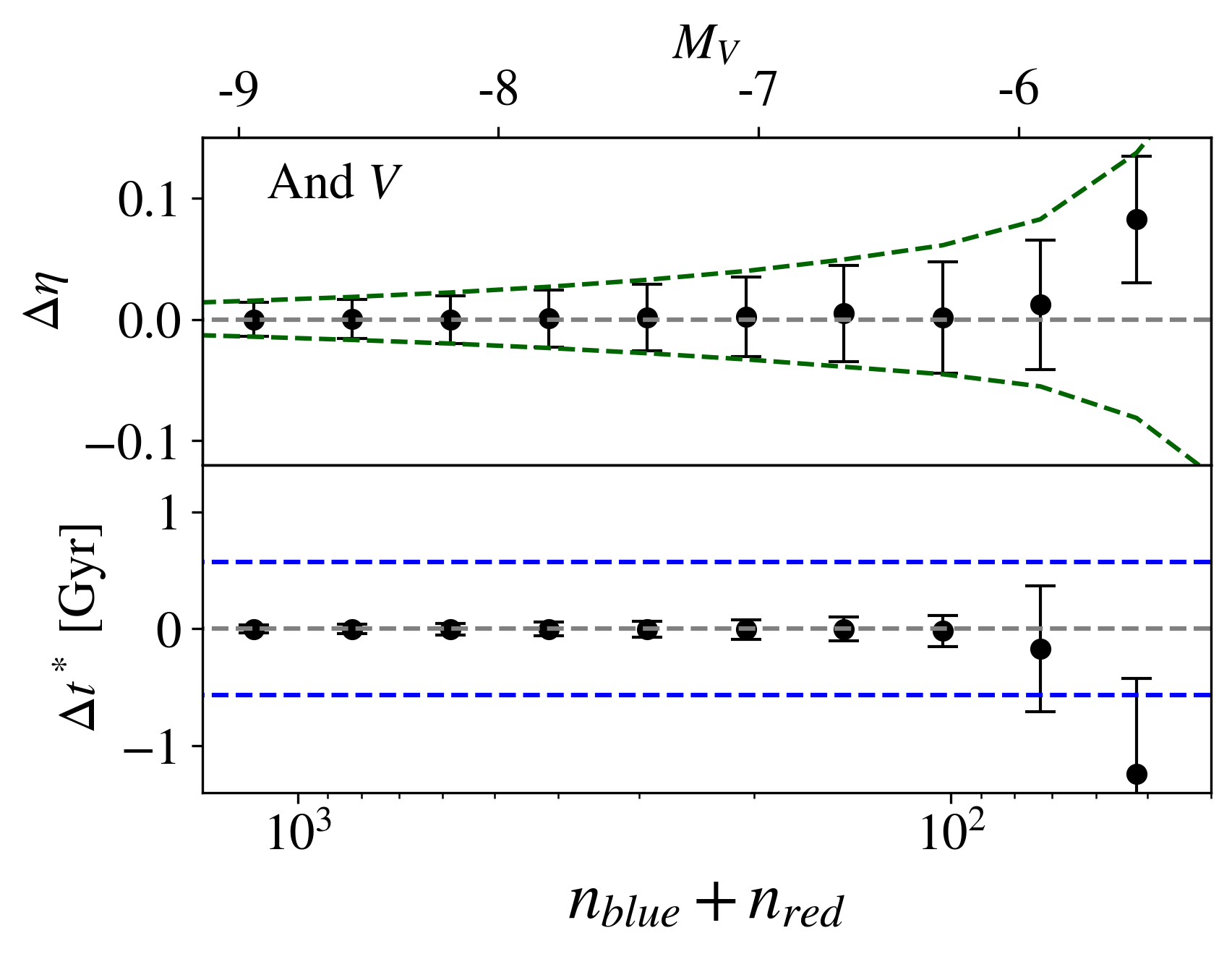} \label{fig: HB population 5}} \quad
    \subfloat
        {\includegraphics[width=0.5\linewidth]{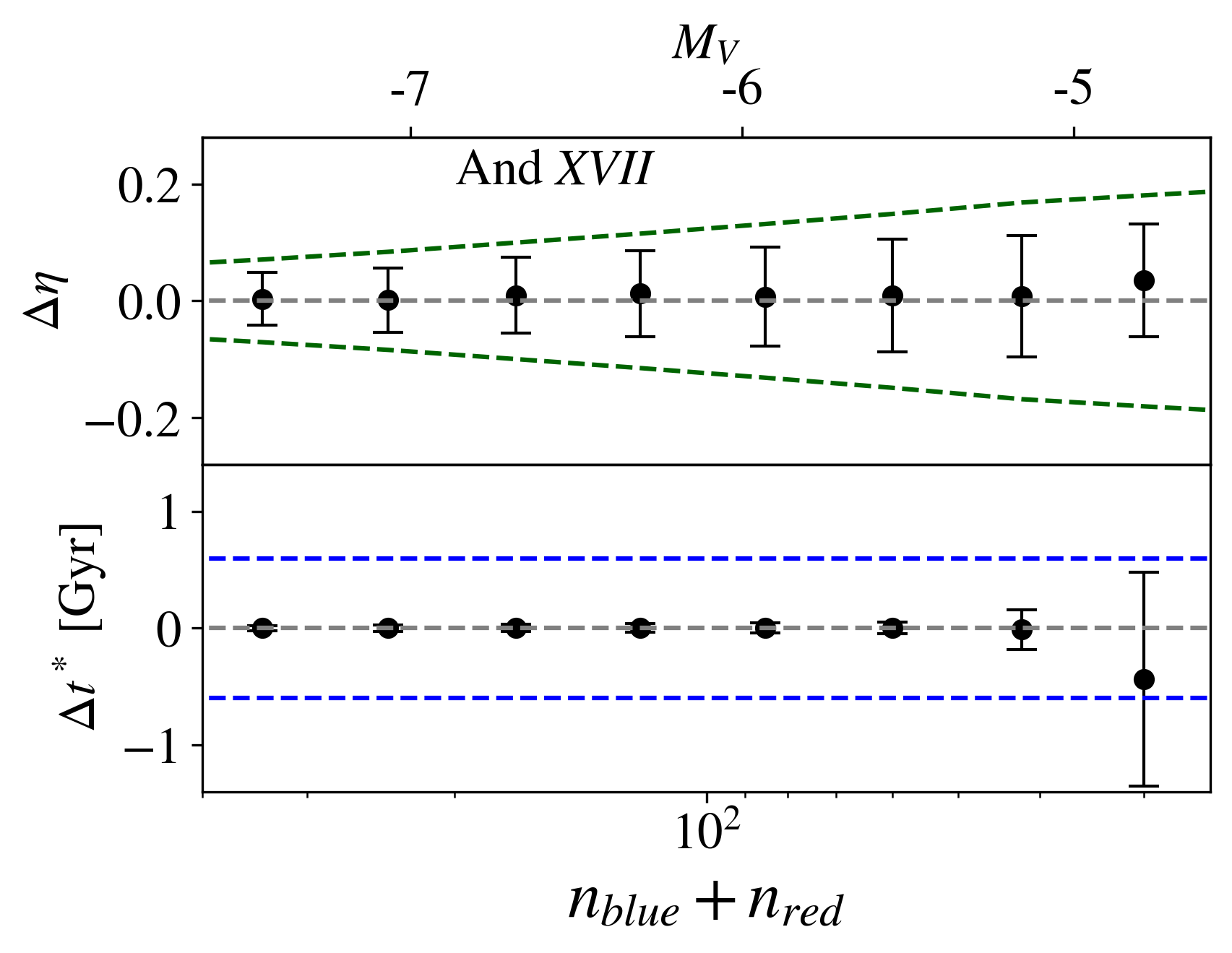} \label{fig: HB population 17}} \quad

    \caption{Differences in the inferred$\eta$ (upper panels) and $t^*$ (lower panels) of \A{V} (left) and \A{XVII} (right) as HB population size is decreased. Black points and error bars show the mean and standard deviation of $\eta$ and $t^*$, calculated from the 1000 Monte Carlo samplings of the HB. Green dashed lines (top panel) show the analytical uncertainty on $\eta$, obtained through the procedure described in section \ref{sec: HB definition}. The blue dashed lines (bottom panel) show the extent of the intrinsic scatter term, from eq~\ref{eq:MCMC fit}. The uncertainty in $t^*$ is dominated by the intrinsic scatter term for $n_{\rm red}+n_{\rm blue}$ $\gtrsim 70$ for \A{V} and $\gtrsim 30$ for \A{XVII}, which corresponds to an absolute magnitude of $M_V \sim -6$ and $M_V \sim -5$, respectively.} %\textbf{For \A{XVII}, the analytical uncertainty of $\eta$ appears to overestimate $\sigma_{\eta}$. This is likely due to how $n_{red}$ and $n_{blue}$ were sampled from the distribution of actual data, which causes the values to be dependent on each other rather than the independent poisson distributions assumed in the analytic uncertainty calculations.}}
    \label{fig: HB population}
 
\end{figure*}

%\begin{figure}
%\centering
%\includegraphics[width=\linewidth]{Figures/HB_population_AND17.png}
%\caption{$\Delta \eta$ and $\Delta t^*$ of \A{XVII} as HB population size is decreased. Construction is the exact same as for \A{V} in figure \ref{fig: HB population}. \A{XVII} has a much larger $\eta$ than \A{V}. Compared to \A{V}, we observe that for galaxies with larger $\eta$ values, increased errors in $\eta$ as $n_{BHB}+n_{RHB}$ decreases doesn't translate to a significant increase in $t^*$ errors until much lower star counts, since the same change in $\eta$ will cause a smaller change in $log(\eta)$.}
%\label{fig: HB population AND17}
%\end{figure}

Figure~\ref{fig: HB population} shows the difference between the original measurements of $\eta$ and $t^*$, and those measured from the re-sampled CMDs, as a function of $n_{HB}$. We do not report any significant bias in the measured ages, even in sparsely populated CMDs. This is likely because stochastic sampling affects stars equally regardless of their position on the HB. The reduced HB population, however, does steadily increase the measurement uncertainty on $\eta$, as it is reasonably expected. Nevertheless, the bottom panels of Fig.~\ref{fig: HB population} show that the contribution of stochastic sampling to the uncertainty on $t^*$ remains subdominant with respect to our calibration scatter for HB populations of $\gtrsim70$ stars for \A{V} and $\gtrsim30$ stars for \A{XVII}. Below these HB population regimes, the low number of HB stars becomes the primary source of uncertainty in age.

The effect of low HB population kicks in at different HB counts for \A{V} (70 stars) and \A{XVII} (30 stars) due to their different HB morphology. Given that eq.~\ref{eq:best-fit} depends on the logarithm of $\eta$, galaxies with blue HB morphologies (i.e., $\eta \sim 1$) will be affected less by a given change in $\eta$ than galaxies with red HBs (i.e., $\eta\sim0$). The corresponding absolute luminosity at which HB sampling becomes relevant is $M_{V} \sim -6$ for \A{V} and $M_{V} \sim -5$ for \A{XVII}. The galaxy luminosity limit for our method applicability is therefore slightly dependent on HB morphology. We note, however, that at such low magnitudes, blue HBs are typically more common.

% \textbf{As $n_{BHB}$ and $n_{RHB}$ approach 0, errors in $t^*$ derived from poisson errors in the $\eta$ calculation can be inaccurate. In particular, at low counts ($\lessapprox 5$) the poisson probability that either $n_{BHB}=0$ or $n_{RHB}=n_{red}-n_{RGB}<0$ becomes significant. These cases lead to non-physical values of $\eta=0$ or $\eta>1$ respectively, that leave the true scatter in $\eta$ and $t^*$ poorly defined. Even if these values of $\eta$ are excluded from the posterior, this can still result in an unrepresentative small set of discrete $\eta$ values that does not represent the full errors in $\eta$ or, once propagated, $t^*$.}

The above experiment assumes that the only source of uncertainty introduced by a low HB population is due to stochastic sampling. In reality, at low HB counts the issue of identifying \textit{bona fide} HB stars become increasingly more relevant. Spurious sources in the HB region, such as photometric artifacts, foreground stars, or background galaxies, can potentially bias the inferred value of $\eta$ and, consequently, of $t^*$. Such an effect is much more difficult to quantify, as it depends on a number of other factors, such as the depth of the photometry, the distance of the galaxy, its galactic latitude, its SFH, and the adopted methodology to reject non-stellar sources. As a general rule, we advise using our calibration with caution whenever the number of  potential contaminants, as estimated by visual inspection of the CMD, is a non-negligible fraction of the stars in the HB region.

\section{Conclusion}
In this paper we explored the connection between the HB morphology of nearby resolved galaxies and their characteristic stellar population ages and metallicities, with the aim of developing a method for estimating the characteristic age of a resolved galaxy outside the LG. 

\vspace{5mm}
1. We use the photometric catalogs and SFHs of 27 M31 satellites developed as part the M31 Satellite Treasury survey to measure a relationship between the ratio of blue-to-red HB stars and the mass-weighted oMSTO age of old/intermediate age stars ($t>6$~Gyr) in the same galaxy.  We find a strong correlation between $\eta$, $t^*$, and $\rm\langle[Fe/H]\rangle$, with older galaxies having, on average, bluer HBs. We derive a linear relationship between these quantities that allows us to estimate the mean-mass weighted age of a stellar population given its HB morphology and luminosity.  The resulting precision is $\lesssim1$~Gyr.  We perform extensive tests to validate these results and explore uncertainties (e.g., filter sets, photometric depth, CMD population).

%The calibration we derive allows us to estimate mean stellar population ages, from the properties of the HB, with precision better than 1~Gyr.

% \vspace{5mm}
% 2. We find three significant outlier to the scaling relation derive above: \A{VI}, \A{XIII}, and \A{XVI}. This is likely connected to the peculiar SFH, and potentially AMR, that these galaxies experienced.

\vspace{5mm}
2. We use this calibration to measure the mass-weighted age of the distant, isolated galaxies KKR~25 and VV~124. For those two galaxies, we measure $t^*$ values of ($11.21^{+0.70}_{-0.65}$~Gyr) and ($11.03^{+0.73}_{-0.68}$~Gyr) for KKR~25 and VV~124, respectively. These values indicate that early star formation in these targets may have proceeded for an extended period of several Gyr. Combined with the current isolated position of these galaxies, this suggests that KKR~25 and VV~124 reached their low specific star formation rate and neutral gas content in isolation, strongly suggesting that internal processes played a major role in their transformation from gas-rich to gas-poor dwarf galaxies.

\vspace{5mm}
3. We characterize the performance of our methodology for a range of photometric depths and galaxy masses. We conclude that our methodology is robust down to SNR of $\sim 10$ on the HB, and for galaxy luminosities as faint as $M_V \sim -5$. This demonstrates that our method can reliably be used to infer characteristic stellar population ages for the hundreds of faint dwarf galaxies that next-generation surveys will discover within a volume of 3-4 Mpc.

\begin{acknowledgements}

Support for this work was provided by NASA through grants GO-13768, GO-15746, GO-15902, AR-16159, and GO-16273 from the Space Telescope Science Institute, which is operated
by AURA, Inc., under NASA contract NAS5-26555.  E.N.K.\ acknowledges support from NSF CAREER grant AST-2233781.  This research has made use of NASA’s Astrophysics Data System Bibliographic Services.

\end{acknowledgements}

\facilities{ \hst\ (ACS)}

\software{ This research made use of routines and modules from the following software packages: \texttt{Astropy} \citep{Astropy}, \texttt{DOLPHOT} \citep{Dolphin16}, \texttt{IPython} \citep{IPython},
\texttt{Matplotlib} \citep{Matplotlib}, \texttt{NumPy} \citep{Numpy}, \texttt{Pandas} \citep{Pandas}, and \texttt{SciPy} \citep{Scipy}.}

\appendix{}
\section{Spectroscopic Metallicites}
\label{App:metallicity}
    In our main analysis, we use metallicity values derived from the local LZ relation \citep{Kirby13b}. This choice was made for the sake of homogeneity and to provide a calibration readily applied to more distant galaxies, which might not have spectroscopic data. Some of our galaxy sample, however, do possess metallicity measurements from RGB spectroscopy.
    In this appendix, we take advantage of these spectroscopic data to quantify systematics due to our LZ-based metallicity assumptions.

\begin{figure}
        \centering
        \includegraphics[width=\linewidth]{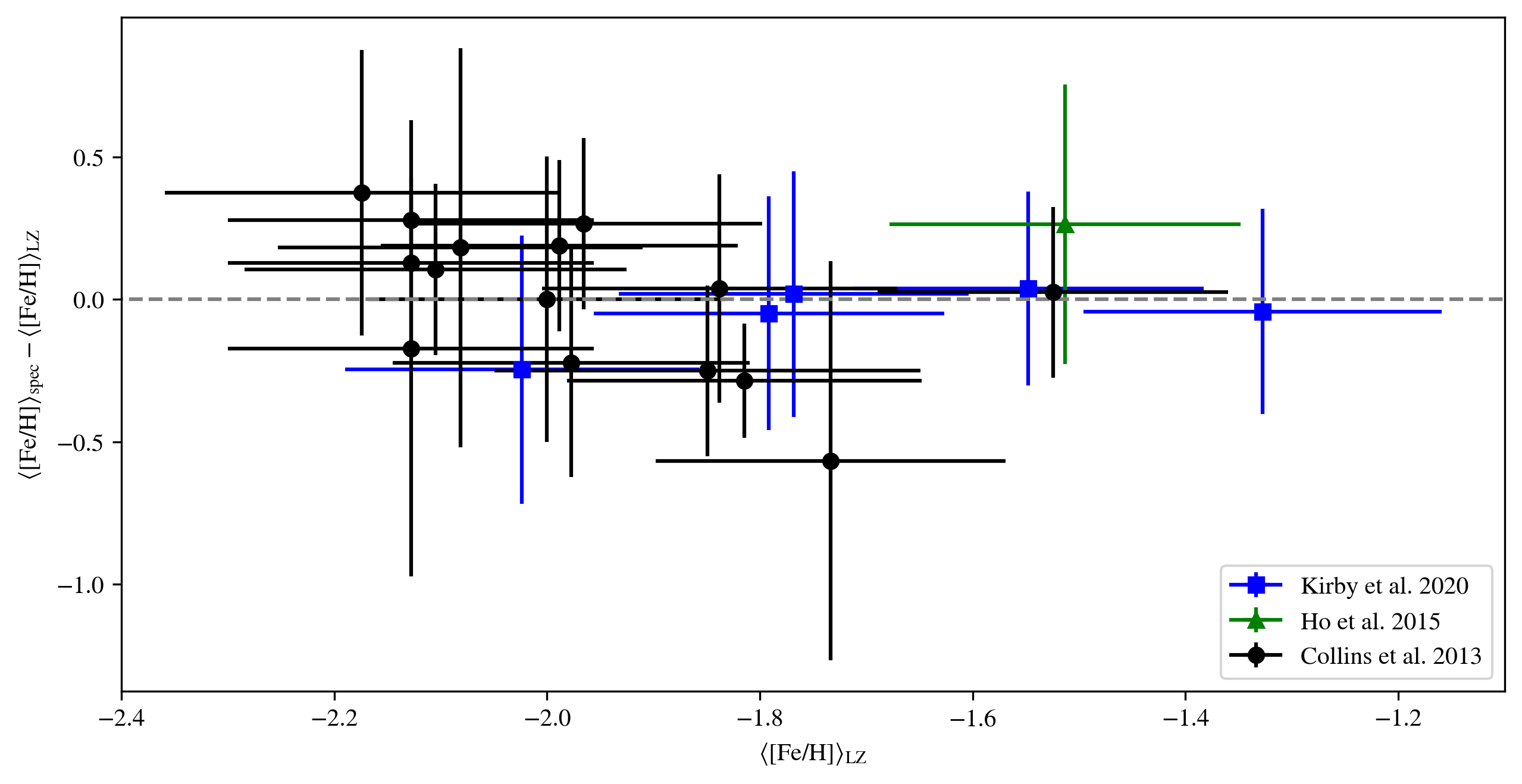}
        \caption{Comparison of mean galaxy metallicities derived from the LZ relation (equation \ref{eq:MetallicityKriby}, taken from \citealt{Kirby2013}) and spectroscopic metalicity values from \citet[blue squares]{Kirby2020}, \citet[green triangle]{Ho2015}, and \citet[black circles]{Collins2013}.}
        \label{fig: LZspec metallicity}
    \end{figure}

    From our sample of 27 galaxies, six have spectroscopic metallicities from a large resolved sample of RGB stars \citep{Vargas14,Ho2015,Kirby2020}. For these, we adopt the $\langle[Fe/H]\rangle$ from \citet{Kirby2020}, with the exception of \A{II}, which we take from \citet{Ho2015}. Furthermore, 15 additional galaxies have $\langle[Fe/H]\rangle$ measurements from stacked RGB spectra \citep{Collins2013}. This gives us a total sample of 21 galaxies with spectroscopic information. Fig.~\ref{fig: LZspec metallicity} shows the comparison between the mean spectroscopic metallicities, $\langle[Fe/H]\rangle_{spec}$, and the values we inferred through eq.~\ref{eq:MetallicityKriby}, $\langle[Fe/H]\rangle_{LZ}$. As demonstrated by the original spectroscopic studies, the two values are in very good agreement, with typical differences comparable to the measurement uncertainties.

\begin{figure}
        \centering
        \includegraphics[width=\linewidth]{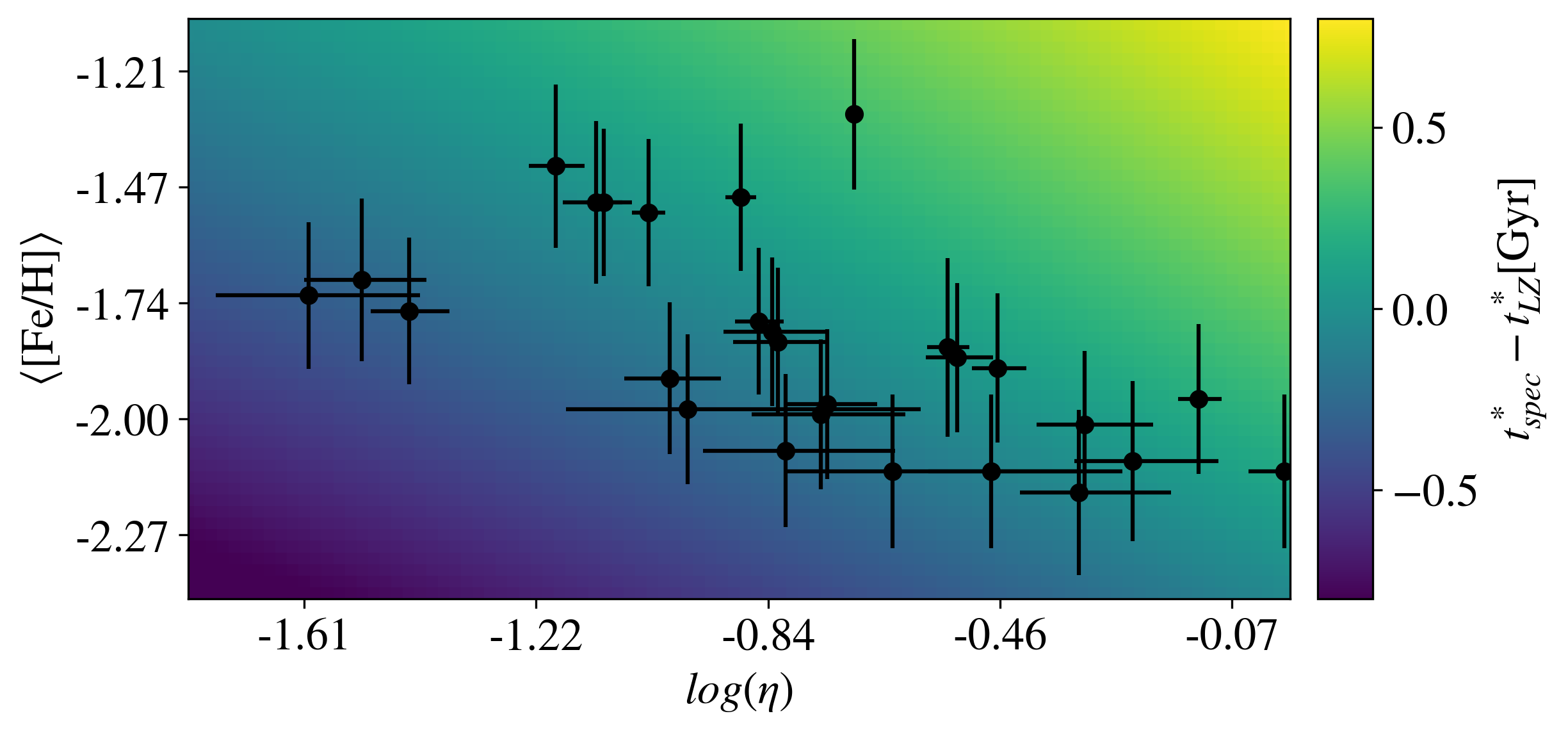}
        \caption{Differences in the measured value of $t^*$, as a function of $\langle[Fe/H]
        \rangle$ and $\eta$, when using the spectroscopy-based calibration, instead of the LZ-based model. The positions of our 27 sample galaxies are shown as black points.}
        \label{fig: LZspec age}
    \end{figure}

    Rerunning our analysis with $\langle[Fe/H]\rangle_{spec}$ yields the set of model coefficients, for eq.~\ref{eq:MCMC fit}, listed in Tab.~\ref{table: LZspec Coef}. While slightly different, this alternative prescription does not result in significantly different values of $t^*$. Fig.~\ref{fig: LZspec age} reports the difference in measured $t^*$, between the LZ-based and spectroscopy-based calibrations, as function of $\eta$ and $\langle[Fe/H]\rangle$. While some differences, upwards of 0.5~Gyr, may arise in some locations of this parameter space, those do not seem to be regions occupied by observed galaxies. In fact, in the $\langle[Fe/H]\rangle$-$\eta$ region spanned by our M31 sample, the mean magnitude $t^*$ difference is 0.12~Gyr and the highest difference is 0.38~Gyr. As an additional test, we have re-measured the ages of our validation galaxies (\S~\ref{sec:validation} and \S~\ref{sec:Recent Star Formation}), as well as those of KKR~25 and VV~124 (\S~\ref{sec:KKVV}). For these galaxies, we estimate $\langle[Fe/H]\rangle$-$\eta$ using the LZ relation, but we calculate $t^*$ using the spectroscopy-based calibration. This emulates a typical use case on more distant galaxies. The resulting $t^*$ values are reported in Tab.~\ref{table: LG galaxies spec}. Also in this case, the inferred ages differ by a small amount, i.e., by less than $\sim$ 200 Myr. The conclusion is that adopting the LZ relation as our calibration baseline does not introduce a significant bias in our analysis.

    \begin{table}
        \caption{Value of the fiducial coefficients from eq.~\ref{eq:MCMC fit}, calibrated against LZ-based metallicities \citep{Kirby2013} and against spectroscopic metallicities from \citet{Collins2013,Ho2015,Kirby2020}.}
        \centering
        \begin{tabular}{ccc}
        
        \toprule
        Name & LZ& Spectroscopy\\
        \toprule
        a & $0.029_{-0.016}^{+0.016}$ & $0.046_{-0.017}^{+0.017}$\\ 
        b & $-0.070_{-0.027}^{+0.028}$ & $-0.046_{-0.020}^{+0.023}$\\ 
        c & $9.967_{-0.059}^{+0.059} $ & $10.023_{-0.041}^{+0.045}$\\ 
        ln(V) & $-7.672_{-0.572}^{+0.447}$ & $-7.808_{-1.141}^{+0.687}$\\
        
        \toprule
        
        \end{tabular}
        \label{table: LZspec Coef}
        
    \end{table}
    
    \begin{table*}
        \caption{Same as Tab.\ref{table: LG galaxies}, but two sets of HB-based $t^*$ are provided, using the LZ-based and Spectroscopy-based calibration, respectively.}
        \centering
        \begin{tabular}{ccccccc}
        
        \toprule
        Name & $\eta$ & $M_{V}$& $\rm \langle[Fe/H]\rangle$ &LZ-based $t^*$ & Spectroscopy-based $t^*$  & $t^*$ from MSTO\\
        &&mag&dex&Gyr&Gyr&Gyr\\
        \toprule
        EriII(F606W) & $0.58^{+0.21}_{-0.13}$ & -7.1 & -2.06±0.19 & $12.74^{+0.83}_{-0.78}(\pm0.86)$ & $12.81^{+0.78}_{-0.74}(\pm0.86)$ & $12.41\pm0.24(\pm1.18)$ \\
        EriII(F475W) & $0.42^{+0.12}_{-0.08}$ & -7.1 & -2.06±0.19 & $12.62^{+0.80}_{-0.75}(\pm0.86)$ & $12.63^{+0.73}_{-0.69}(\pm0.86)$ & $12.41\pm0.24(\pm1.18)$ \\
        Cetus & $0.07^{+0.01}_{-0.01}$ & -11.2 & -1.58±0.18 & $11.09^{+0.69}_{-0.65}(\pm0.86)$ & $11.05^{+0.64}_{-0.61}(\pm0.86)$ & $10.75\pm0.08(\pm0.67)$ \\
        Tucana & $0.43^{+0.03}_{-0.02}$ & -9.5 & -1.78±0.18 & $12.07^{+0.77}_{-0.72}(\pm0.86)$ & $12.27^{+0.69}_{-0.65}(\pm0.86)$ & $12.06\pm0.12(\pm0.74)$ \\
        Pisces & $0.09^{+0.03}_{-0.03}$ & -9.8 & -1.74±0.18 & $11.48^{+0.70}_{-0.66}(\pm0.86)$ & $11.39^{+0.64}_{-0.60}(\pm0.86)$ & $10.87\pm0.19(\pm0.54)$ \\
        PegDIG & $0.01^{+0.01}_{-0.01}$ & -12.3 & -1.45±0.18 & $10.31^{+0.74}_{-0.69}(\pm0.86)$ & $10.01^{+0.76}_{-0.71}(\pm0.86)$ & $9.81^{+0.28}_{-0.30}(\pm1.31)$ \\
        KKR25 & $0.06^{+0.01}_{-0.01}$ & -10.5 & -1.66±0.18 & $11.21^{+0.70}_{-0.65}(\pm0.86)$ & $11.10^{+0.64}_{-0.61}(\pm0.86)$ & - \\
        VV124 & $0.13^{+0.01}_{-0.01}$ & -12.5 & -1.43±0.18 & $11.03^{+0.73}_{-0.68}(\pm0.86)$ & $11.20^{+0.65}_{-0.61}(\pm0.86)$ & - \\
        \toprule
        
        \end{tabular}
        \label{table: LG galaxies spec}  
    
    \end{table*}

\bibliography{sample631}{}

\begin{thebibliography}{}
\expandafter\ifx\csname natexlab\endcsname\relax\def\natexlab#1{#1}\fi
\providecommand{\url}[1]{\href{#1}{#1}}
\providecommand{\dodoi}[1]{doi:~\href{http://doi.org/#1}{\nolinkurl{#1}}}
\providecommand{\doeprint}[1]{\href{http://ascl.net/#1}{\nolinkurl{http://ascl.net/#1}}}
\providecommand{\doarXiv}[1]{\href{https://arxiv.org/abs/#1}{\nolinkurl{https://arxiv.org/abs/#1}}}

\bibitem[{{Astropy Collaboration} {et~al.}(2013){Astropy Collaboration},
  {Robitaille}, {Tollerud}, {Greenfield}, {Droettboom}, {Bray}, {Aldcroft},
  {Davis}, {Ginsburg}, {Price-Whelan}, {Kerzendorf}, {Conley}, {Crighton},
  {Barbary}, {Muna}, {Ferguson}, {Grollier}, {Parikh}, {Nair}, {Unther},
  {Deil}, {Woillez}, {Conseil}, {Kramer}, {Turner}, {Singer}, {Fox}, {Weaver},
  {Zabalza}, {Edwards}, {Azalee Bostroem}, {Burke}, {Casey}, {Crawford},
  {Dencheva}, {Ely}, {Jenness}, {Labrie}, {Lim}, {Pierfederici}, {Pontzen},
  {Ptak}, {Refsdal}, {Servillat}, \& {Streicher}}]{Astropy}
{Astropy Collaboration}, {Robitaille}, T.~P., {Tollerud}, E.~J., {et~al.} 2013,
  \aap, 558, A33, \dodoi{10.1051/0004-6361/201322068}

\bibitem[{{Bastian} \& {Lardo}(2018)}]{Bastian18}
{Bastian}, N., \& {Lardo}, C. 2018, \araa, 56, 83,
  \dodoi{10.1146/annurev-astro-081817-051839}

\bibitem[{{Begum} \& {Chengalur}(2005)}]{Begum05}
{Begum}, A., \& {Chengalur}, J.~N. 2005, \mnras, 362, 609,
  \dodoi{10.1111/j.1365-2966.2005.09342.x}

\bibitem[{{Bellazzini} {et~al.}(2011{\natexlab{a}}){Bellazzini}, {Perina},
  {Galleti}, \& {Oosterloo}}]{Bellazzini11b}
{Bellazzini}, M., {Perina}, S., {Galleti}, S., \& {Oosterloo}, T.
  2011{\natexlab{a}}, \aap, 533, A37, \dodoi{10.1051/0004-6361/201117275}

\bibitem[{{Bellazzini} {et~al.}(2011{\natexlab{b}}){Bellazzini}, {Beccari},
  {Oosterloo}, {Galleti}, {Sollima}, {Correnti}, {Testa}, {Mayer}, {Cignoni},
  {Fraternali}, \& {Gallozzi}}]{Bellazzini11a}
{Bellazzini}, M., {Beccari}, G., {Oosterloo}, T.~A., {et~al.}
  2011{\natexlab{b}}, \aap, 527, A58, \dodoi{10.1051/0004-6361/201016159}

\bibitem[{{Brown} {et~al.}(2014){Brown}, {Tumlinson}, {Geha}, {Simon},
  {Vargas}, {VandenBerg}, {Kirby}, {Kalirai}, {Avila}, {Gennaro}, {Ferguson},
  {Mu{\~n}oz}, {Guhathakurta}, \& {Renzini}}]{Brown14}
{Brown}, T.~M., {Tumlinson}, J., {Geha}, M., {et~al.} 2014, \apj, 796, 91,
  \dodoi{10.1088/0004-637X/796/2/91}

\bibitem[{{Cassisi} {et~al.}(2013){Cassisi}, {Mucciarelli}, {Pietrinferni},
  {Salaris}, \& {Ferguson}}]{Cassisi13}
{Cassisi}, S., {Mucciarelli}, A., {Pietrinferni}, A., {Salaris}, M., \&
  {Ferguson}, J. 2013, \aap, 554, A19, \dodoi{10.1051/0004-6361/201321311}

\bibitem[{{Catelan} {et~al.}(2009){Catelan}, {Grundahl}, {Sweigart},
  {Valcarce}, \& {Cort{\'e}s}}]{Catelan09}
{Catelan}, M., {Grundahl}, F., {Sweigart}, A.~V., {Valcarce}, A.~A.~R., \&
  {Cort{\'e}s}, C. 2009, \apjl, 695, L97, \dodoi{10.1088/0004-637X/695/1/L97}

\bibitem[{{Cignoni} \& {Tosi}(2010)}]{Cignoni10}
{Cignoni}, M., \& {Tosi}, M. 2010, Advances in Astronomy, 2010, 158568,
  \dodoi{10.1155/2010/158568}

\bibitem[{{Clementini} {et~al.}(2005){Clementini}, {Ripepi}, {Bragaglia},
  {Martinez Fiorenzano}, {Held}, \& {Gratton}}]{Clementini05}
{Clementini}, G., {Ripepi}, V., {Bragaglia}, A., {et~al.} 2005, \mnras, 363,
  734, \dodoi{10.1111/j.1365-2966.2005.09478.x}

\bibitem[{{Cole} {et~al.}(2014){Cole}, {Weisz}, {Dolphin}, {Skillman},
  {McConnachie}, {Brooks}, \& {Leaman}}]{Cole14}
{Cole}, A.~A., {Weisz}, D.~R., {Dolphin}, A.~E., {et~al.} 2014, \apj, 795, 54,
  \dodoi{10.1088/0004-637X/795/1/54}

\bibitem[{{Cole} {et~al.}(2007){Cole}, {Skillman}, {Tolstoy}, {Gallagher},
  {Aparicio}, {Dolphin}, {Gallart}, {Hidalgo}, {Saha}, {Stetson}, \&
  {Weisz}}]{Cole07}
{Cole}, A.~A., {Skillman}, E.~D., {Tolstoy}, E., {et~al.} 2007, \apjl, 659,
  L17, \dodoi{10.1086/516711}

\bibitem[{{Collins} {et~al.}(2013){Collins}, {Chapman}, {Rich}, {Ibata},
  {Martin}, {Irwin}, {Bate}, {Lewis}, {Pe{\~n}arrubia}, {Arimoto}, {Casey},
  {Ferguson}, {Koch}, {McConnachie}, \& {Tanvir}}]{Collins2013}
{Collins}, M. L.~M., {Chapman}, S.~C., {Rich}, R.~M., {et~al.} 2013, \apj, 768,
  172, \dodoi{10.1088/0004-637X/768/2/172}

\bibitem[{{Crnojevi{\'c}} {et~al.}(2016){Crnojevi{\'c}}, {Sand}, {Zaritsky},
  {Spekkens}, {Willman}, \& {Hargis}}]{Crnojevic16}
{Crnojevi{\'c}}, D., {Sand}, D.~J., {Zaritsky}, D., {et~al.} 2016, \apjl, 824,
  L14, \dodoi{10.3847/2041-8205/824/1/L14}

\bibitem[{{Da Costa} {et~al.}(2002){Da Costa}, {Armandroff}, \&
  {Caldwell}}]{Dacosta02}
{Da Costa}, G.~S., {Armandroff}, T.~E., \& {Caldwell}, N. 2002, \aj, 124, 332,
  \dodoi{10.1086/340965}

\bibitem[{{Da Costa} {et~al.}(1996){Da Costa}, {Armandroff}, {Caldwell}, \&
  {Seitzer}}]{Dacosta96}
{Da Costa}, G.~S., {Armandroff}, T.~E., {Caldwell}, N., \& {Seitzer}, P. 1996,
  \aj, 112, 2576, \dodoi{10.1086/118204}

\bibitem[{{Da Costa} {et~al.}(2000){Da Costa}, {Armandroff}, {Caldwell}, \&
  {Seitzer}}]{Dacosta00}
---. 2000, \aj, 119, 705, \dodoi{10.1086/301223}

\bibitem[{{Dalcanton} {et~al.}(2009){Dalcanton}, {Williams}, {Seth}, {Dolphin},
  {Holtzman}, {Rosema}, {Skillman}, {Cole}, {Girardi}, {Gogarten},
  {Karachentsev}, {Olsen}, {Weisz}, {Christensen}, {Freeman}, {Gilbert},
  {Gallart}, {Harris}, {Hodge}, {de Jong}, {Karachentseva}, {Mateo}, {Stetson},
  {Tavarez}, {Zaritsky}, {Governato}, \& {Quinn}}]{Dalcanton09}
{Dalcanton}, J.~J., {Williams}, B.~F., {Seth}, A.~C., {et~al.} 2009, \apjs,
  183, 67, \dodoi{10.1088/0067-0049/183/1/67}

\bibitem[{{Dalessandro} {et~al.}(2011){Dalessandro}, {Salaris}, {Ferraro},
  {Cassisi}, {Lanzoni}, {Rood}, {Fusi Pecci}, \& {Sabbi}}]{Dalessandro11}
{Dalessandro}, E., {Salaris}, M., {Ferraro}, F.~R., {et~al.} 2011, \mnras, 410,
  694, \dodoi{10.1111/j.1365-2966.2010.17479.x}

\bibitem[{{de Boer} {et~al.}(2014){de Boer}, {Tolstoy}, {Lemasle}, {Saha},
  {Olszewski}, {Mateo}, {Irwin}, \& {Battaglia}}]{deBoer14}
{de Boer}, T.~J.~L., {Tolstoy}, E., {Lemasle}, B., {et~al.} 2014, \aap, 572,
  A10, \dodoi{10.1051/0004-6361/201424119}

\bibitem[{{de Boer} {et~al.}(2012{\natexlab{a}}){de Boer}, {Tolstoy}, {Hill},
  {Saha}, {Olsen}, {Starkenburg}, {Lemasle}, {Irwin}, \&
  {Battaglia}}]{deBoer12a}
{de Boer}, T.~J.~L., {Tolstoy}, E., {Hill}, V., {et~al.} 2012{\natexlab{a}},
  \aap, 539, A103, \dodoi{10.1051/0004-6361/201118378}

\bibitem[{{de Boer} {et~al.}(2012{\natexlab{b}}){de Boer}, {Tolstoy}, {Hill},
  {Saha}, {Olszewski}, {Mateo}, {Starkenburg}, {Battaglia}, \&
  {Walker}}]{deBoer12b}
---. 2012{\natexlab{b}}, \aap, 544, A73, \dodoi{10.1051/0004-6361/201219547}

\bibitem[{{Dolphin}(2016)}]{Dolphin16}
{Dolphin}, A. 2016, {DOLPHOT: Stellar photometry}, Astrophysics Source Code
  Library, record ascl:1608.013.
\newblock \doeprint{1608.013}

\bibitem[{{Dolphin}(2000)}]{Dolphin00}
{Dolphin}, A.~E. 2000, \pasp, 112, 1383, \dodoi{10.1086/316630}

\bibitem[{{Dolphin}(2002)}]{Dolphin02}
---. 2002, \mnras, 332, 91, \dodoi{10.1046/j.1365-8711.2002.05271.x}

\bibitem[{{Dolphin}(2012)}]{Dolphin12}
---. 2012, \apj, 751, 60, \dodoi{10.1088/0004-637X/751/1/60}

\bibitem[{{Dolphin}(2013)}]{Dolphin13}
---. 2013, \apj, 775, 76, \dodoi{10.1088/0004-637X/775/1/76}

\bibitem[{{Dotter} {et~al.}(2010){Dotter}, {Sarajedini}, {Anderson},
  {Aparicio}, {Bedin}, {Chaboyer}, {Majewski}, {Mar{\'\i}n-Franch}, {Milone},
  {Paust}, {Piotto}, {Reid}, {Rosenberg}, \& {Siegel}}]{Dotter10}
{Dotter}, A., {Sarajedini}, A., {Anderson}, J., {et~al.} 2010, \apj, 708, 698,
  \dodoi{10.1088/0004-637X/708/1/698}

\bibitem[{{Fabrizio} {et~al.}(2015){Fabrizio}, {Nonino}, {Bono}, {Primas},
  {Th{\'e}venin}, {Stetson}, {Cassisi}, {Buonanno}, {Coppola}, {da Silva},
  {Dall'Ora}, {Ferraro}, {Genovali}, {Gilmozzi}, {Iannicola}, {Marconi},
  {Monelli}, {Romaniello}, \& {Walker}}]{Fabrizio15}
{Fabrizio}, M., {Nonino}, M., {Bono}, G., {et~al.} 2015, \aap, 580, A18,
  \dodoi{10.1051/0004-6361/201525753}

\bibitem[{{Foreman-Mackey} {et~al.}(2013){Foreman-Mackey}, {Hogg}, {Lang}, \&
  {Goodman}}]{Foreman_Mackey_2013_MCMC}
{Foreman-Mackey}, D., {Hogg}, D.~W., {Lang}, D., \& {Goodman}, J. 2013, \pasp,
  125, 306, \dodoi{10.1086/670067}

\bibitem[{{Fu} {et~al.}(2022){Fu}, {Weisz}, {Starkenburg}, {Martin}, {Ji},
  {Patel}, {Boylan-Kolchin}, {C{\^o}t{\'e}}, {Dolphin}, {Longeard}, {Mateo}, \&
  {Sandford}}]{Fu22}
{Fu}, S.~W., {Weisz}, D.~R., {Starkenburg}, E., {et~al.} 2022, \apj, 925, 6,
  \dodoi{10.3847/1538-4357/ac3665}

\bibitem[{{Gallart} {et~al.}(2005){Gallart}, {Zoccali}, \&
  {Aparicio}}]{Gallart05}
{Gallart}, C., {Zoccali}, M., \& {Aparicio}, A. 2005, \araa, 43, 387,
  \dodoi{10.1146/annurev.astro.43.072103.150608}

\bibitem[{{Gallart} {et~al.}(2015){Gallart}, {Monelli}, {Mayer}, {Aparicio},
  {Battaglia}, {Bernard}, {Cassisi}, {Cole}, {Dolphin}, {Drozdovsky},
  {Hidalgo}, {Navarro}, {Salvadori}, {Skillman}, {Stetson}, \&
  {Weisz}}]{Gallart15}
{Gallart}, C., {Monelli}, M., {Mayer}, L., {et~al.} 2015, \apjl, 811, L18,
  \dodoi{10.1088/2041-8205/811/2/L18}

\bibitem[{{Gallart} {et~al.}(2021){Gallart}, {Monelli}, {Ruiz-Lara},
  {Calamida}, {Cassisi}, {Cignoni}, {Anderson}, {Battaglia}, {Bermejo-Climent},
  {Bernard}, {Mart{\'\i}nez-V{\'a}zquez}, {Mayer}, {Salvadori}, {Monachesi},
  {Navarro}, {Shen}, {Surot}, {Tosi}, {Bajaj}, \& {Strinfellow}}]{Gallart21}
{Gallart}, C., {Monelli}, M., {Ruiz-Lara}, T., {et~al.} 2021, \apj, 909, 192,
  \dodoi{10.3847/1538-4357/abddbe}

\bibitem[{{Geha} {et~al.}(2012){Geha}, {Blanton}, {Yan}, \& {Tinker}}]{Geha12}
{Geha}, M., {Blanton}, M.~R., {Yan}, R., \& {Tinker}, J.~L. 2012, \apj, 757,
  85, \dodoi{10.1088/0004-637X/757/1/85}

\bibitem[{{Geisler} {et~al.}(2007){Geisler}, {Wallerstein}, {Smith}, \&
  {Casetti-Dinescu}}]{Geisler07}
{Geisler}, D., {Wallerstein}, G., {Smith}, V.~V., \& {Casetti-Dinescu}, D.~I.
  2007, \pasp, 119, 939, \dodoi{10.1086/521990}

\bibitem[{{Gratton} {et~al.}(2010){Gratton}, {Carretta}, {Bragaglia},
  {Lucatello}, \& {D'Orazi}}]{Gratton10}
{Gratton}, R.~G., {Carretta}, E., {Bragaglia}, A., {Lucatello}, S., \&
  {D'Orazi}, V. 2010, \aap, 517, A81, \dodoi{10.1051/0004-6361/200912572}

\bibitem[{{Gratton} {et~al.}(2011){Gratton}, {Lucatello}, {Carretta},
  {Bragaglia}, {D'Orazi}, \& {Momany}}]{Gratton11}
{Gratton}, R.~G., {Lucatello}, S., {Carretta}, E., {et~al.} 2011, \aap, 534,
  A123, \dodoi{10.1051/0004-6361/201117690}

\bibitem[{{Grebel} \& {Gallagher}(2004)}]{Grebel04}
{Grebel}, E.~K., \& {Gallagher}, John~S., I. 2004, \apjl, 610, L89,
  \dodoi{10.1086/423339}

\bibitem[{{Grebel} {et~al.}(2003){Grebel}, {Gallagher}, \&
  {Harbeck}}]{Grebel03}
{Grebel}, E.~K., {Gallagher}, John~S., I., \& {Harbeck}, D. 2003, \aj, 125,
  1926, \dodoi{10.1086/368363}

\bibitem[{{Green} {et~al.}(2019){Green}, {Schlafly}, {Zucker}, {Speagle}, \&
  {Finkbeiner}}]{Green19}
{Green}, G.~M., {Schlafly}, E., {Zucker}, C., {Speagle}, J.~S., \&
  {Finkbeiner}, D. 2019, \apj, 887, 93, \dodoi{10.3847/1538-4357/ab5362}

\bibitem[{{Harbeck} {et~al.}(2001){Harbeck}, {Grebel}, {Holtzman},
  {Guhathakurta}, {Brandner}, {Geisler}, {Sarajedini}, {Dolphin},
  {Hurley-Keller}, \& {Mateo}}]{Harbeck01}
{Harbeck}, D., {Grebel}, E.~K., {Holtzman}, J., {et~al.} 2001, \aj, 122, 3092,
  \dodoi{10.1086/324232}

\bibitem[{{Hidalgo} {et~al.}(2011){Hidalgo}, {Aparicio}, {Skillman}, {Monelli},
  {Gallart}, {Cole}, {Dolphin}, {Weisz}, {Bernard}, {Cassisi}, {Mayer},
  {Stetson}, {Tolstoy}, \& {Ferguson}}]{Hidalgo11}
{Hidalgo}, S.~L., {Aparicio}, A., {Skillman}, E., {et~al.} 2011, \apj, 730, 14,
  \dodoi{10.1088/0004-637X/730/1/14}

\bibitem[{{Hidalgo} {et~al.}(2018){Hidalgo}, {Pietrinferni}, {Cassisi},
  {Salaris}, {Mucciarelli}, {Savino}, {Aparicio}, {Silva Aguirre}, \&
  {Verma}}]{Hidalgo18}
{Hidalgo}, S.~L., {Pietrinferni}, A., {Cassisi}, S., {et~al.} 2018, \apj, 856,
  125, \dodoi{10.3847/1538-4357/aab158}

\bibitem[{{Ho} {et~al.}(2015){Ho}, {Geha}, {Tollerud}, {Zinn}, {Guhathakurta},
  \& {Vargas}}]{Ho2015}
{Ho}, N., {Geha}, M., {Tollerud}, E.~J., {et~al.} 2015, \apj, 798, 77,
  \dodoi{10.1088/0004-637X/798/2/77}

\bibitem[{{Hunter}(2007)}]{Matplotlib}
{Hunter}, J.~D. 2007, Computing in Science Engineering, 9, 90

\bibitem[{{Karachentsev} {et~al.}(2004){Karachentsev}, {Karachentseva},
  {Huchtmeier}, \& {Makarov}}]{Karachentsev04}
{Karachentsev}, I.~D., {Karachentseva}, V.~E., {Huchtmeier}, W.~K., \&
  {Makarov}, D.~I. 2004, \aj, 127, 2031, \dodoi{10.1086/382905}

\bibitem[{{Karachentsev} {et~al.}(2013){Karachentsev}, {Makarov}, \&
  {Kaisina}}]{Karachentsev13}
{Karachentsev}, I.~D., {Makarov}, D.~I., \& {Kaisina}, E.~I. 2013, \aj, 145,
  101, \dodoi{10.1088/0004-6256/145/4/101}

\bibitem[{{Karachentsev} {et~al.}(2001){Karachentsev}, {Sharina}, {Dolphin},
  {Geisler}, {Grebel}, {Guhathakurta}, {Hodge}, {Karachentseva}, {Sarajedini},
  \& {Seitzer}}]{Karachentsev01}
{Karachentsev}, I.~D., {Sharina}, M.~E., {Dolphin}, A.~E., {et~al.} 2001, \aap,
  379, 407, \dodoi{10.1051/0004-6361:20011344}

\bibitem[{{Kirby} {et~al.}(2012){Kirby}, {Cohen}, \& {Bellazzini}}]{Kirby12}
{Kirby}, E.~N., {Cohen}, J.~G., \& {Bellazzini}, M. 2012, \apj, 751, 46,
  \dodoi{10.1088/0004-637X/751/1/46}

\bibitem[{{Kirby} {et~al.}(2013{\natexlab{a}}){Kirby}, {Cohen}, \&
  {Bellazzini}}]{Kirby13b}
---. 2013{\natexlab{a}}, \apj, 768, 96, \dodoi{10.1088/0004-637X/768/1/96}

\bibitem[{{Kirby} {et~al.}(2013{\natexlab{b}}){Kirby}, {Cohen}, {Guhathakurta},
  {Cheng}, {Bullock}, \& {Gallazzi}}]{Kirby2013}
{Kirby}, E.~N., {Cohen}, J.~G., {Guhathakurta}, P., {et~al.}
  2013{\natexlab{b}}, \apj, 779, 102, \dodoi{10.1088/0004-637X/779/2/102}

\bibitem[{{Kirby} {et~al.}(2020){Kirby}, {Gilbert}, {Escala}, {Wojno},
  {Guhathakurta}, {Majewski}, \& {Beaton}}]{Kirby2020}
{Kirby}, E.~N., {Gilbert}, K.~M., {Escala}, I., {et~al.} 2020, \aj, 159, 46,
  \dodoi{10.3847/1538-3881/ab5f0f}

\bibitem[{{Kopylov} {et~al.}(2008){Kopylov}, {Tikhonov}, {Fabrika},
  {Drozdovsky}, \& {Valeev}}]{Kopylov08}
{Kopylov}, A.~I., {Tikhonov}, N.~A., {Fabrika}, S., {Drozdovsky}, I., \&
  {Valeev}, A.~F. 2008, \mnras, 387, L45,
  \dodoi{10.1111/j.1745-3933.2008.00482.x}

\bibitem[{{Lee} {et~al.}(2009){Lee}, {Yuk}, {Park}, {Harris}, \&
  {Zaritsky}}]{Lee09}
{Lee}, M.~G., {Yuk}, I.-S., {Park}, H.~S., {Harris}, J., \& {Zaritsky}, D.
  2009, \apj, 703, 692, \dodoi{10.1088/0004-637X/703/1/692}

\bibitem[{{Li} {et~al.}(2017){Li}, {Simon}, {Drlica-Wagner}, {Bechtol}, {Wang},
  {Garc{\'\i}a-Bellido}, {Frieman}, {Marshall}, {James}, {Strigari}, {Pace},
  {Balbinot}, {Zhang}, {Abbott}, {Allam}, {Benoit-L{\'e}vy}, {Bernstein},
  {Bertin}, {Brooks}, {Burke}, {Carnero Rosell}, {Carrasco Kind}, {Carretero},
  {Cunha}, {D'Andrea}, {da Costa}, {DePoy}, {Desai}, {Diehl}, {Eifler},
  {Flaugher}, {Goldstein}, {Gruen}, {Gruendl}, {Gschwend}, {Gutierrez},
  {Krause}, {Kuehn}, {Lin}, {Maia}, {March}, {Menanteau}, {Miquel}, {Plazas},
  {Romer}, {Sanchez}, {Santiago}, {Schubnell}, {Sevilla-Noarbe}, {Smith},
  {Sobreira}, {Suchyta}, {Tarle}, {Thomas}, {Tucker}, {Walker}, {Wechsler},
  {Wester}, {Yanny}, \& {DES Collaboration}}]{Li17}
{Li}, T.~S., {Simon}, J.~D., {Drlica-Wagner}, A., {et~al.} 2017, \apj, 838, 8,
  \dodoi{10.3847/1538-4357/aa6113}

\bibitem[{{Mackey} \& {van den Bergh}(2005)}]{Mackey05}
{Mackey}, A.~D., \& {van den Bergh}, S. 2005, \mnras, 360, 631,
  \dodoi{10.1111/j.1365-2966.2005.09080.x}

\bibitem[{{Makarov} {et~al.}(2012){Makarov}, {Makarova}, {Sharina}, {Uklein},
  {Tikhonov}, {Guhathakurta}, {Kirby}, \& {Terekhova}}]{Makarov12}
{Makarov}, D., {Makarova}, L., {Sharina}, M., {et~al.} 2012, \mnras, 425, 709,
  \dodoi{10.1111/j.1365-2966.2012.21581.x}

\bibitem[{{Martell} {et~al.}(2011){Martell}, {Smolinski}, {Beers}, \&
  {Grebel}}]{Martell11}
{Martell}, S.~L., {Smolinski}, J.~P., {Beers}, T.~C., \& {Grebel}, E.~K. 2011,
  \aap, 534, A136, \dodoi{10.1051/0004-6361/201117644}

\bibitem[{{Martin} {et~al.}(2017){Martin}, {Weisz}, {Albers}, {Bernard},
  {Collins}, {Dolphin}, {Ferguson}, {Ibata}, {Laevens}, {Lewis}, {Mackey},
  {McConnachie}, {Rich}, \& {Skillman}}]{2017Martin}
{Martin}, N.~F., {Weisz}, D.~R., {Albers}, S.~M., {et~al.} 2017, \apj, 850, 16,
  \dodoi{10.3847/1538-4357/aa901a}

\bibitem[{{Mart{\'\i}nez-V{\'a}zquez}
  {et~al.}(2021){Mart{\'\i}nez-V{\'a}zquez}, {Monelli}, {Cassisi}, {Taibi},
  {Gallart}, {Vivas}, {Walker}, {Mart{\'\i}n-Ravelo}, {Zenteno}, {Battaglia},
  {Bono}, {Calamida}, {Carollo}, {Cicu{\'e}ndez}, {Fiorentino}, {Marconi},
  {Salvadori}, {Balbinot}, {Bernard}, {Dall'Ora}, \&
  {Stetson}}]{MartinezVazquez21}
{Mart{\'\i}nez-V{\'a}zquez}, C.~E., {Monelli}, M., {Cassisi}, S., {et~al.}
  2021, \mnras, 508, 1064, \dodoi{10.1093/mnras/stab2493}

\bibitem[{{Mateo}(1998)}]{Mateo98}
{Mateo}, M.~L. 1998, \araa, 36, 435, \dodoi{10.1146/annurev.astro.36.1.435}

\bibitem[{{McConnachie}(2012)}]{McConnachie12}
{McConnachie}, A.~W. 2012, \aj, 144, 4, \dodoi{10.1088/0004-6256/144/1/4}

\bibitem[{{McQuinn} {et~al.}(2010){McQuinn}, {Skillman}, {Cannon}, {Dalcanton},
  {Dolphin}, {Hidalgo-Rodr{\'\i}guez}, {Holtzman}, {Stark}, {Weisz}, \&
  {Williams}}]{McQuinn10}
{McQuinn}, K. B.~W., {Skillman}, E.~D., {Cannon}, J.~M., {et~al.} 2010, \apj,
  721, 297, \dodoi{10.1088/0004-637X/721/1/297}

\bibitem[{{Milone} {et~al.}(2014){Milone}, {Marino}, {Dotter}, {Norris},
  {Jerjen}, {Piotto}, {Cassisi}, {Bedin}, {Recio Blanco}, {Sarajedini},
  {Asplund}, {Monelli}, \& {Aparicio}}]{Milone14}
{Milone}, A.~P., {Marino}, A.~F., {Dotter}, A., {et~al.} 2014, \apj, 785, 21,
  \dodoi{10.1088/0004-637X/785/1/21}

\bibitem[{{Milone} {et~al.}(2018){Milone}, {Marino}, {Renzini}, {D'Antona},
  {Anderson}, {Barbuy}, {Bedin}, {Bellini}, {Brown}, {Cassisi}, {Cordoni},
  {Lagioia}, {Nardiello}, {Ortolani}, {Piotto}, {Sarajedini}, {Tailo}, {van der
  Marel}, \& {Vesperini}}]{Milone18}
{Milone}, A.~P., {Marino}, A.~F., {Renzini}, A., {et~al.} 2018, \mnras, 481,
  5098, \dodoi{10.1093/mnras/sty2573}

\bibitem[{{Monelli} {et~al.}(2010){Monelli}, {Hidalgo}, {Stetson}, {Aparicio},
  {Gallart}, {Dolphin}, {Cole}, {Weisz}, {Skillman}, {Bernard}, {Mayer},
  {Navarro}, {Cassisi}, {Drozdovsky}, \& {Tolstoy}}]{Monelli10}
{Monelli}, M., {Hidalgo}, S.~L., {Stetson}, P.~B., {et~al.} 2010, \apj, 720,
  1225, \dodoi{10.1088/0004-637X/720/2/1225}

\bibitem[{{Monelli} {et~al.}(2016){Monelli}, {Mart{\'\i}nez-V{\'a}zquez},
  {Bernard}, {Gallart}, {Skillman}, {Weisz}, {Dolphin}, {Hidalgo}, {Cole},
  {Martin}, {Aparicio}, {Cassisi}, {Boylan-Kolchin}, {Mayer}, {McConnachie},
  {McQuinn}, \& {Navarro}}]{Monelli16}
{Monelli}, M., {Mart{\'\i}nez-V{\'a}zquez}, C.~E., {Bernard}, E.~J., {et~al.}
  2016, \apj, 819, 147, \dodoi{10.3847/0004-637X/819/2/147}

\bibitem[{{Mu{\~n}oz} {et~al.}(2018{\natexlab{a}}){Mu{\~n}oz}, {C{\^o}t{\'e}},
  {Santana}, {Geha}, {Simon}, {Oyarz{\'u}n}, {Stetson}, \&
  {Djorgovski}}]{Munoz18a}
{Mu{\~n}oz}, R.~R., {C{\^o}t{\'e}}, P., {Santana}, F.~A., {et~al.}
  2018{\natexlab{a}}, \apj, 860, 65, \dodoi{10.3847/1538-4357/aac168}

\bibitem[{{Mu{\~n}oz} {et~al.}(2018{\natexlab{b}}){Mu{\~n}oz}, {C{\^o}t{\'e}},
  {Santana}, {Geha}, {Simon}, {Oyarz{\'u}n}, {Stetson}, \&
  {Djorgovski}}]{Munoz18b}
---. 2018{\natexlab{b}}, \apj, 860, 66, \dodoi{10.3847/1538-4357/aac16b}

\bibitem[{{Mutlu-Pakdil} {et~al.}(2021){Mutlu-Pakdil}, {Sand}, {Crnojevi{\'c}},
  {Drlica-Wagner}, {Caldwell}, {Guhathakurta}, {Seth}, {Simon}, {Strader}, \&
  {Toloba}}]{Mutlu-Pakdil21}
{Mutlu-Pakdil}, B., {Sand}, D.~J., {Crnojevi{\'c}}, D., {et~al.} 2021, \apj,
  918, 88, \dodoi{10.3847/1538-4357/ac0db8}

\bibitem[{{Nagarajan} {et~al.}(2022){Nagarajan}, {Weisz}, \&
  {El-Badry}}]{Nagarajan22}
{Nagarajan}, P., {Weisz}, D.~R., \& {El-Badry}, K. 2022, \apj, 932, 19,
  \dodoi{10.3847/1538-4357/ac69e6}

\bibitem[{{Neeley} {et~al.}(2021){Neeley}, {Monelli}, {Marengo}, {Fiorentino},
  {Vivas}, {Walker}, {Gallart}, {Mart{\'\i}nez-V{\'a}zquez}, {Bono}, {Cassisi},
  {Marconi}, {Dall'Ora}, \& {Sarajedini}}]{Neeley21}
{Neeley}, J.~R., {Monelli}, M., {Marengo}, M., {et~al.} 2021, \apj, 920, 152,
  \dodoi{10.3847/1538-4357/ac1a7a}

\bibitem[{{Perez} \& {Granger}(2007)}]{IPython}
{Perez}, F., \& {Granger}, B.~E. 2007, Computing in Science Engineering, 9, 21

\bibitem[{{Piotto} {et~al.}(2007){Piotto}, {Bedin}, {Anderson}, {King},
  {Cassisi}, {Milone}, {Villanova}, {Pietrinferni}, \& {Renzini}}]{Piotto07}
{Piotto}, G., {Bedin}, L.~R., {Anderson}, J., {et~al.} 2007, \apjl, 661, L53,
  \dodoi{10.1086/518503}

\bibitem[{{Putman} {et~al.}(2021){Putman}, {Zheng}, {Price-Whelan}, {Grcevich},
  {Johnson}, {Tollerud}, \& {Peek}}]{Putman21}
{Putman}, M.~E., {Zheng}, Y., {Price-Whelan}, A.~M., {et~al.} 2021, \apj, 913,
  53, \dodoi{10.3847/1538-4357/abe391}

\bibitem[{{Qu} {et~al.}(2023){Qu}, {Yuan}, {Doliva-Dolinsky}, {Martin}, {Kang},
  {Wei}, {Li}, {Luo}, {Chang}, {Tsai}, {Fan}, \& {Ibata}}]{Qu23}
{Qu}, H., {Yuan}, Z., {Doliva-Dolinsky}, A., {et~al.} 2023, \mnras, 523, 876,
  \dodoi{10.1093/mnras/stad1352}

\bibitem[{{Rejkuba} {et~al.}(2005){Rejkuba}, {Greggio}, {Harris}, {Harris}, \&
  {Peng}}]{Rejkuba05}
{Rejkuba}, M., {Greggio}, L., {Harris}, W.~E., {Harris}, G. L.~H., \& {Peng},
  E.~W. 2005, \apj, 631, 262, \dodoi{10.1086/432462}

\bibitem[{{Rejkuba} {et~al.}(2011){Rejkuba}, {Harris}, {Greggio}, \&
  {Harris}}]{Rejkuba11}
{Rejkuba}, M., {Harris}, W.~E., {Greggio}, L., \& {Harris}, G.~L.~H. 2011,
  \aap, 526, A123, \dodoi{10.1051/0004-6361/201015640}

\bibitem[{{Rose} {et~al.}(2005){Rose}, {Arimoto}, {Caldwell}, {Schiavon},
  {Vazdekis}, \& {Yamada}}]{Rose05}
{Rose}, J.~A., {Arimoto}, N., {Caldwell}, N., {et~al.} 2005, \aj, 129, 712,
  \dodoi{10.1086/427136}

\bibitem[{{Rusakov} {et~al.}(2021){Rusakov}, {Monelli}, {Gallart}, {Fritz},
  {Ruiz-Lara}, {Bernard}, \& {Cassisi}}]{Rusakov21}
{Rusakov}, V., {Monelli}, M., {Gallart}, C., {et~al.} 2021, \mnras, 502, 642,
  \dodoi{10.1093/mnras/stab006}

\bibitem[{{Salaris} {et~al.}(2013){Salaris}, {de Boer}, {Tolstoy},
  {Fiorentino}, \& {Cassisi}}]{Salaris13}
{Salaris}, M., {de Boer}, T., {Tolstoy}, E., {Fiorentino}, G., \& {Cassisi}, S.
  2013, \aap, 559, A57, \dodoi{10.1051/0004-6361/201322501}

\bibitem[{{Sales} {et~al.}(2007){Sales}, {Navarro}, {Abadi}, \&
  {Steinmetz}}]{Sales07}
{Sales}, L.~V., {Navarro}, J.~F., {Abadi}, M.~G., \& {Steinmetz}, M. 2007,
  \mnras, 379, 1475, \dodoi{10.1111/j.1365-2966.2007.12026.x}

\bibitem[{{Sandage} \& {Wildey}(1967)}]{Sandage67}
{Sandage}, A., \& {Wildey}, R. 1967, \apj, 150, 469, \dodoi{10.1086/149350}

\bibitem[{{Sarajedini} {et~al.}(1995){Sarajedini}, {Lee}, \&
  {Lee}}]{Sarajedini95}
{Sarajedini}, A., {Lee}, Y.-W., \& {Lee}, D.-H. 1995, \apj, 450, 712,
  \dodoi{10.1086/176177}

\bibitem[{{Sarajedini} {et~al.}(2007){Sarajedini}, {Bedin}, {Chaboyer},
  {Dotter}, {Siegel}, {Anderson}, {Aparicio}, {King}, {Majewski},
  {Mar{\'\i}n-Franch}, {Piotto}, {Reid}, \& {Rosenberg}}]{Sarajedini07}
{Sarajedini}, A., {Bedin}, L.~R., {Chaboyer}, B., {et~al.} 2007, \aj, 133,
  1658, \dodoi{10.1086/511979}

\bibitem[{{Savino} {et~al.}(2018){Savino}, {de Boer}, {Salaris}, \&
  {Tolstoy}}]{Savino18}
{Savino}, A., {de Boer}, T.~J.~L., {Salaris}, M., \& {Tolstoy}, E. 2018,
  \mnras, 480, 1587, \dodoi{10.1093/mnras/sty1954}

\bibitem[{{Savino} {et~al.}(2020){Savino}, {Koch}, {Prudil}, {Kunder}, \&
  {Smolec}}]{Savino20}
{Savino}, A., {Koch}, A., {Prudil}, Z., {Kunder}, A., \& {Smolec}, R. 2020,
  \aap, 641, A96, \dodoi{10.1051/0004-6361/202038305}

\bibitem[{{Savino} {et~al.}(2019){Savino}, {Tolstoy}, {Salaris}, {Monelli}, \&
  {de Boer}}]{Savino19}
{Savino}, A., {Tolstoy}, E., {Salaris}, M., {Monelli}, M., \& {de Boer},
  T.~J.~L. 2019, \aap, 630, A116, \dodoi{10.1051/0004-6361/201936077}

\bibitem[{{Savino} {et~al.}(2022){Savino}, {Weisz}, {Skillman}, {Dolphin},
  {Kallivayalil}, {Wetzel}, {Anderson}, {Besla}, {Boylan-Kolchin}, {Bullock},
  {Cole}, {Collins}, {Cooper}, {Deason}, {Dotter}, {Fardal}, {Ferguson},
  {Fritz}, {Geha}, {Gilbert}, {Guhathakurta}, {Ibata}, {Irwin}, {Jeon},
  {Kirby}, {Lewis}, {Mackey}, {Majewski}, {Martin}, {McConnachie}, {Patel},
  {Rich}, {Simon}, {Sohn}, {Tollerud}, \& {van der Marel}}]{Savino22}
{Savino}, A., {Weisz}, D.~R., {Skillman}, E.~D., {et~al.} 2022, \apj, 938, 101,
  \dodoi{10.3847/1538-4357/ac91cb}

\bibitem[{{Savino} {et~al.}(2023){Savino}, {Weisz}, {Skillman}, {Dolphin},
  {Cole}, {Kallivayalil}, {Wetzel}, {Anderson}, {Besla}, {Boylan-Kolchin},
  {Brown}, {Bullock}, {Collins}, {Cooper}, {Deason}, {Dotter}, {Fardal},
  {Ferguson}, {Fritz}, {Geha}, {Gilbert}, {Guhathakurta}, {Ibata}, {Irwin},
  {Jeon}, {Kirby}, {Lewis}, {Mackey}, {Majewski}, {Martin}, {McConnachie},
  {Patel}, {Rich}, {Simon}, {Sohn}, {Tollerud}, \& {van der Marel}}]{Savino23}
---. 2023, arXiv e-prints, arXiv:2305.13360, \dodoi{10.48550/arXiv.2305.13360}

\bibitem[{{Schiavon} {et~al.}(2017){Schiavon}, {Johnson}, {Frinchaboy},
  {Zasowski}, {M{\'e}sz{\'a}ros}, {Garc{\'\i}a-Hern{\'a}ndez}, {Cohen}, {Tang},
  {Villanova}, {Geisler}, {Beers}, {Fern{\'a}ndez-Trincado}, {Garc{\'\i}a
  P{\'e}rez}, {Lucatello}, {Majewski}, {Martell}, {O'Connell}, {Allende
  Prieto}, {Bizyaev}, {Carrera}, {Lane}, {Malanushenko}, {Malanushenko},
  {Mu{\~n}oz}, {Nitschelm}, {Oravetz}, {Pan}, {Roman-Lopes}, {Schultheis}, \&
  {Simmons}}]{Schiavon17}
{Schiavon}, R.~P., {Johnson}, J.~A., {Frinchaboy}, P.~M., {et~al.} 2017,
  \mnras, 466, 1010, \dodoi{10.1093/mnras/stw3093}

\bibitem[{{Simon}(2019)}]{Simon19}
{Simon}, J.~D. 2019, \araa, 57, 375,
  \dodoi{10.1146/annurev-astro-091918-104453}

\bibitem[{{Simon} {et~al.}(2021){Simon}, {Brown}, {Drlica-Wagner}, {Li},
  {Avila}, {Bechtol}, {Clementini}, {Crnojevi{\'c}}, {Garofalo}, {Geha},
  {Sand}, {Strader}, \& {Willman}}]{Simon21}
{Simon}, J.~D., {Brown}, T.~M., {Drlica-Wagner}, A., {et~al.} 2021, \apj, 908,
  18, \dodoi{10.3847/1538-4357/abd31b}

\bibitem[{{Skillman} {et~al.}(2017){Skillman}, {Monelli}, {Weisz}, {Hidalgo},
  {Aparicio}, {Bernard}, {Boylan-Kolchin}, {Cassisi}, {Cole}, {Dolphin},
  {Ferguson}, {Gallart}, {Irwin}, {Martin}, {Mart{\'\i}nez-V{\'a}zquez},
  {Mayer}, {McConnachie}, {McQuinn}, {Navarro}, \& {Stetson}}]{Skillman17}
{Skillman}, E.~D., {Monelli}, M., {Weisz}, D.~R., {et~al.} 2017, \apj, 837,
  102, \dodoi{10.3847/1538-4357/aa60c5}

\bibitem[{{Taibi} {et~al.}(2022){Taibi}, {Battaglia}, {Leaman}, {Brooks},
  {Riggs}, {Munshi}, {Revaz}, \& {Jablonka}}]{Taibi22}
{Taibi}, S., {Battaglia}, G., {Leaman}, R., {et~al.} 2022, \aap, 665, A92,
  \dodoi{10.1051/0004-6361/202243508}

\bibitem[{{Taibi} {et~al.}(2020){Taibi}, {Battaglia}, {Rejkuba}, {Leaman},
  {Kacharov}, {Iorio}, {Jablonka}, \& {Zoccali}}]{Taibi20}
{Taibi}, S., {Battaglia}, G., {Rejkuba}, M., {et~al.} 2020, \aap, 635, A152,
  \dodoi{10.1051/0004-6361/201937240}

\bibitem[{{Taibi} {et~al.}(2018){Taibi}, {Battaglia}, {Kacharov}, {Rejkuba},
  {Irwin}, {Leaman}, {Zoccali}, {Tolstoy}, \& {Jablonka}}]{Taibi18}
{Taibi}, S., {Battaglia}, G., {Kacharov}, N., {et~al.} 2018, \aap, 618, A122,
  \dodoi{10.1051/0004-6361/201833414}

\bibitem[{{Tailo} {et~al.}(2020){Tailo}, {Milone}, {Lagioia}, {D'Antona},
  {Marino}, {Vesperini}, {Caloi}, {Ventura}, {Dondoglio}, \&
  {Cordoni}}]{Tailo20}
{Tailo}, M., {Milone}, A.~P., {Lagioia}, E.~P., {et~al.} 2020, \mnras, 498,
  5745, \dodoi{10.1093/mnras/staa2639}

\bibitem[{{Teyssier} {et~al.}(2012){Teyssier}, {Johnston}, \&
  {Kuhlen}}]{Teyssier12}
{Teyssier}, M., {Johnston}, K.~V., \& {Kuhlen}, M. 2012, \mnras, 426, 1808,
  \dodoi{10.1111/j.1365-2966.2012.21793.x}

\bibitem[{{Tolstoy} {et~al.}(2009){Tolstoy}, {Hill}, \& {Tosi}}]{Tolstoy09}
{Tolstoy}, E., {Hill}, V., \& {Tosi}, M. 2009, \araa, 47, 371,
  \dodoi{10.1146/annurev-astro-082708-101650}

\bibitem[{{Tully} {et~al.}(2013){Tully}, {Courtois}, {Dolphin}, {Fisher},
  {H{\'e}raudeau}, {Jacobs}, {Karachentsev}, {Makarov}, {Makarova},
  {Mitronova}, {Rizzi}, {Shaya}, {Sorce}, \& {Wu}}]{Tully13}
{Tully}, R.~B., {Courtois}, H.~M., {Dolphin}, A.~E., {et~al.} 2013, \aj, 146,
  86, \dodoi{10.1088/0004-6256/146/4/86}

\bibitem[{{van den Bergh}(1967)}]{vandenBergh67}
{van den Bergh}, S. 1967, \aj, 72, 70, \dodoi{10.1086/110203}

\bibitem[{{van der Walt} {et~al.}(2011){van der Walt}, {Colbert}, \&
  {Varoquaux}}]{Numpy}
{van der Walt}, S., {Colbert}, S.~C., \& {Varoquaux}, G. 2011, Computing in
  Science Engineering, 13, 22

\bibitem[{{VandenBerg} {et~al.}(2013){VandenBerg}, {Brogaard}, {Leaman}, \&
  {Casagrande}}]{Vandenberg13}
{VandenBerg}, D.~A., {Brogaard}, K., {Leaman}, R., \& {Casagrande}, L. 2013,
  \apj, 775, 134, \dodoi{10.1088/0004-637X/775/2/134}

\bibitem[{{Vargas} {et~al.}(2014){Vargas}, {Geha}, \& {Tollerud}}]{Vargas14}
{Vargas}, L.~C., {Geha}, M.~C., \& {Tollerud}, E.~J. 2014, \apj, 790, 73,
  \dodoi{10.1088/0004-637X/790/1/73}

\bibitem[{{Virtanen} {et~al.}(2020){Virtanen}, {Gommers}, {Oliphant},
  {Haberland}, {Reddy}, {Cournapeau}, {Burovski}, {Peterson}, {Weckesser},
  {Bright}, {van der Walt}, {Brett}, {Wilson}, {Jarrod Millman}, {Mayorov},
  {Nelson}, {Jones}, {Kern}, {Larson}, {Carey}, {Polat}, {Feng}, {Moore}, {Vand
  erPlas}, {Laxalde}, {Perktold}, {Cimrman}, {Henriksen}, {Quintero}, {Harris},
  {Archibald}, {Ribeiro}, {Pedregosa}, {van Mulbregt}, \&
  {Contributors}}]{Scipy}
{Virtanen}, P., {Gommers}, R., {Oliphant}, T.~E., {et~al.} 2020, Nature
  Methods, 17, 261, \dodoi{https://doi.org/10.1038/s41592-019-0686-2}

\bibitem[{{Weisz} {et~al.}(2014{\natexlab{a}}){Weisz}, {Dolphin}, {Skillman},
  {Holtzman}, {Gilbert}, {Dalcanton}, \& {Williams}}]{Weisz14}
{Weisz}, D.~R., {Dolphin}, A.~E., {Skillman}, E.~D., {et~al.}
  2014{\natexlab{a}}, \apj, 789, 148, \dodoi{10.1088/0004-637X/789/2/148}

\bibitem[{{Weisz} {et~al.}(2011){Weisz}, {Dalcanton}, {Williams}, {Gilbert},
  {Skillman}, {Seth}, {Dolphin}, {McQuinn}, {Gogarten}, {Holtzman}, {Rosema},
  {Cole}, {Karachentsev}, \& {Zaritsky}}]{Weisz11}
{Weisz}, D.~R., {Dalcanton}, J.~J., {Williams}, B.~F., {et~al.} 2011, \apj,
  739, 5, \dodoi{10.1088/0004-637X/739/1/5}

\bibitem[{{Weisz} {et~al.}(2014{\natexlab{b}}){Weisz}, {Skillman}, {Hidalgo},
  {Monelli}, {Dolphin}, {McConnachie}, {Bernard}, {Gallart}, {Aparicio},
  {Boylan-Kolchin}, {Cassisi}, {Cole}, {Ferguson}, {Irwin}, {Martin}, {Mayer},
  {McQuinn}, {Navarro}, \& {Stetson}}]{Weisz14c}
{Weisz}, D.~R., {Skillman}, E.~D., {Hidalgo}, S.~L., {et~al.}
  2014{\natexlab{b}}, \apj, 789, 24, \dodoi{10.1088/0004-637X/789/1/24}

\bibitem[{{Weisz} {et~al.}(2015){Weisz}, {Johnson}, {Foreman-Mackey},
  {Dolphin}, {Beerman}, {Williams}, {Dalcanton}, {Rix}, {Hogg}, {Fouesneau},
  {Johnson}, {Bell}, {Boyer}, {Gouliermis}, {Guhathakurta}, {Kalirai}, {Lewis},
  {Seth}, \& {Skillman}}]{Weisz15}
{Weisz}, D.~R., {Johnson}, L.~C., {Foreman-Mackey}, D., {et~al.} 2015, \apj,
  806, 198, \dodoi{10.1088/0004-637X/806/2/198}

\bibitem[{{W}es {M}c{K}inney(2010)}]{Pandas}
{W}es {M}c{K}inney. 2010, in {P}roceedings of the 9th {P}ython in {S}cience
  {C}onference, ed. {S}t\'efan van~der {W}alt \& {J}arrod {M}illman, 56 -- 61,
  \dodoi{10.25080/Majora-92bf1922-00a}

\end{thebibliography}
\bibliographystyle{aasjournal}



\end{document}